\documentclass{emulateapj}
\usepackage{natbib}
\usepackage{amssymb}
\usepackage{setspace}
\usepackage{graphicx}
\usepackage{rotating}
\submitted{Accepted for publication in the Astrophysical Journal}

\def\CIVdblt{{\rm C~}\kern 0.1em{\sc iv}~$\lambda\lambda 1548, 1550$}
\def\MgIIdblt{{\rm Mg~}\kern 0.1em{\sc ii}~$\lambda\lambda 2796, 2803$}
\def\NVdblt{{\rm N}\kern 0.1em{\sc v}~$\lambda\lambda 1238, 1242$}  
\def\OVIdblt{{\rm O}\kern 0.1em{\sc vi}~$\lambda\lambda 1031, 1037$}
\def\SiIVdblt{{\rm Si~}\kern 0.1em{\sc iv}~$\lambda\lambda1394, 1403$}
\def\AlIIIdblt{{\rm Al~}\kern 0.1em{\sc iii}~$\lambda\lambda1855,1863$}
\def\FeIIdblt{{\rm Fe~}\kern 0.1em{\sc ii}~$\lambda\lambda 2383, 2600$}
\def\NeVIIIdblt{{\rm Ne~}\kern 0.1em{\sc viii}~$\lambda\lambda 770, 780$}

\def\NeVIII{\hbox{{\rm Ne~}\kern 0.1em{\sc viii}}}
\def\OI{\hbox{{\rm O~}\kern 0.1em{\sc i}}}
\def\OII{\hbox{{\rm O~}\kern 0.1em{\sc ii}}}
\def\OIII{\hbox{{\rm O~}\kern 0.1em{\sc iii}}}
\def\OIV{\hbox{{\rm O~}\kern 0.1em{\sc iv}}}
\def\OVI{\hbox{{\rm O~}\kern 0.1em{\sc vi}}}
\def\OVII{\hbox{{\rm O~}\kern 0.1em{\sc vii}}}
\def\OVIII{\hbox{{\rm O~}\kern 0.1em{\sc viii}}}
\def\NIII{\hbox{{\rm N~}\kern 0.1em{\sc iii}}}
\def\NIV{\hbox{{\rm N~}\kern 0.1em{\sc iv}}}
\def\NVII{\hbox{{\rm N~}\kern 0.1em{\sc vii}}}
\def\CIII{\hbox{{\rm C~}\kern 0.1em{\sc iii}}}
\def\SiIII{\hbox{{\rm Si~}\kern 0.1em{\sc iii}}}
\def\SVI{\hbox{{\rm S~}\kern 0.1em{\sc vi}}}
\def\NeIX{\hbox{{\rm Ne~}\kern 0.1em{\sc ix}}}

\def\AlII{\hbox{{\rm Al~}\kern 0.1em{\sc ii}}}
\def\AlIII{\hbox{{\rm Al~}\kern 0.1em{\sc iii}}}
\def\CaI{\hbox{{\rm Ca}\kern 0.1em{\sc i}}}
\def\CaII{\hbox{{\rm Ca}\kern 0.1em{\sc ii}}}
\def\CrII{\hbox{{\rm Cr}\kern 0.1em{\sc ii}}}
\def\CII{\hbox{{\rm C~}\kern 0.1em{\sc ii}}}
\def\CIII{\hbox{{\rm C~}\kern 0.1em{\sc iii}}}
\def\CIV{\hbox{{\rm C~}\kern 0.1em{\sc iv}}}
\def\CV{\hbox{{\rm C}\kern 0.1em{\sc v}}}
\def\H{\hbox{{\rm H}}}
\def\HI{\hbox{{\rm H~}\kern 0.1em{\sc i}}}
\def\HII{\hbox{{\rm H~}\kern 0.1em{\sc ii}}}
\def\Lya{\hbox{{\rm Ly}\kern 0.1em$\alpha$}}
\def\Lyb{\hbox{{\rm Ly}\kern 0.1em$\beta$}}
\def\Lyg{\hbox{{\rm Ly}\kern 0.1em$\gamma$}}
\def\Lyth{\hbox{{\rm Ly}\kern 0.1em$\theta$}}
\def\Lyfive{\hbox{{\rm Ly}\kern 0.1em$5$}}
\def\Lysix{\hbox{{\rm Ly}\kern 0.1em$6$}}
\def\Lyseven{\hbox{{\rm Ly}\kern 0.1em$7$}}
\def\Lyeight{\hbox{{\rm Ly}\kern 0.1em$8$}}
\def\Lynine{\hbox{{\rm Ly}\kern 0.1em$9$}}
\def\Lyten{\hbox{{\rm Ly}\kern 0.1em$10$}}
\def\HeI{\hbox{{\rm He}\kern 0.1em{\sc i}}}
\def\HeII{\hbox{{\rm He}\kern 0.1em{\sc ii}}}
\def\FeI{\hbox{{\rm Fe~}\kern 0.1em{\sc i}}}
\def\FeII{\hbox{{\rm Fe~}\kern 0.1em{\sc ii}}}
\def\FeIII{\hbox{{\rm Fe~}\kern 0.1em{\sc iii}}}
\def\MnII{\hbox{{\rm Mn}\kern 0.1em{\sc ii}}}
\def\MgI{\hbox{{\rm Mg~}\kern 0.1em{\sc i}}}
\def\MgII{\hbox{{\rm Mg~}\kern 0.1em{\sc ii}}}
\def\MgIII{\hbox{{\rm Mg~}\kern 0.1em{\sc iii}}}
\def\MgIV{\hbox{{\rm Mg~}\kern 0.1em{\sc iv}}}
\def\NaI{\hbox{{\rm Na}\kern 0.1em{\sc i}}}
\def\NV{\hbox{{\rm N}\kern 0.1em{\sc v}}}
\def\NII{\hbox{{\rm N}\kern 0.1em{\sc ii}}}
\def\NIII{\hbox{{\rm N}\kern 0.1em{\sc iii}}}
\def\OVI{\hbox{{\rm O}\kern 0.1em{\sc vi}}}
\def\SiII{\hbox{{\rm Si~}\kern 0.1em{\sc ii}}}
\def\SiIII{\hbox{{\rm Si~}\kern 0.1em{\sc iii}}}
\def\SiIV{\hbox{{\rm Si~}\kern 0.1em{\sc iv}}}
\def\SII{\hbox{{\rm S}\kern 0.1em{\sc ii}}}
\def\SIII{\hbox{{\rm S}\kern 0.1em{\sc iii}}}
\def\SIV{\hbox{{\rm S}\kern 0.1em{\sc iv}}}
\def\TiII{\hbox{{\rm Ti}\kern 0.1em{\sc ii}}}
\def\ZnII{\hbox{{\rm Zn}\kern 0.1em{\sc ii}}}
\def\ArI{\hbox{{\rm Ar}\kern 0.1em{\sc i}}}
\newcommand{\kms}{\hbox{km~s$^{-1}$}}
\newcommand{\cmsq}{\hbox{cm$^{-2}$}}
\newcommand{\cc}{\hbox{cm$^{-3}$}}
\def\kms{\hbox{km~s$^{-1}$}}      
\def\cmsq{\hbox{cm$^{-2}$}}
\def\cc{\hbox{cm$^{-3}$}}
\def\etal{et~al.\ }

\begin{document}

\title{Cosmic Origins Spectrograph and $FUSE$ Observations \\ of $T \sim 10^5$~K Gas In A Nearby Galaxy Filament\altaffilmark{1}} 
\author{Anand Narayanan, Bart P. Wakker, Blair D. Savage\altaffilmark{2}, Brian A. Keeney,  \\ J. Michael Shull, John T. Stocke\altaffilmark{3} \& Kenneth R. Sembach\altaffilmark{4}}

\altaffiltext{1}{Based on observations with the NASA/ESA {\it Hubble Space Telescope}, obtained at the Space Telescope Science Institute, which is operated by the Association of Universities for Research in Astronomy, Inc., under NASA contract NAS 05-26555, and the NASA-CNES/ESA {\it Far Ultraviolet SpectroscopicExplorer} mission, operated by the Johns Hopkins University, supported by NASA contract NAS 05-32985.}
\altaffiltext{2}{Department of Astronomy, The University of Wisconsin-Madison, 5534 Sterling Hall, 475 N. Charter Street, Madison WI 53706-1582, USA, Email: anand, wakker, wakker@astro.wisc.edu}
\altaffiltext{3}{CASA, Department of Astrophysical and Planetary Sciences, University of Colorado, 389-UCB, Boulder, CO 80309, USA.}
\altaffiltext{4}{Space Telescope Science Institute, 3700 San Martin Drive, Baltimore, MD 21218, USA.}

\begin{abstract}

We present a clear detection of a broad {\Lya} absorber (BLA) with a matching {\OVI} line in the nearby universe. The BLA is detected at $z(\Lya) = 0.01028$ in the high $S/N$ spectrum of Mrk~290 obtained using the Cosmic Origins Spectrograph. The {\Lya} absorption has two components, with $b(\HI) = 55~{\pm}~1$~{\kms} and $b(\HI) = 33~{\pm}~1$~{\kms}, separated in velocity by $v \sim 115$~{\kms}. The {\OVI}, detected by FUSE at $z(\OVI) = 0.01027$, has a $b(\OVI) = 29~{\pm}~3$~{\kms} and is kinematically well aligned with the broader {\HI} component. The non-detection of other ions such as {\CII}, {\SiII}, {\FeII}, {\CIII}, {\SiIII}, {\CIV}, {\SiIV} and {\NV} at the same velocity as the BLA and the {\OVI} implies that the absorber is tracing highly ionized gas. The different line widths of the BLA and {\OVI} suggest a temperature of $T = 1.4 \times 10^5$~K in the absorber. Photoionziation, collisional ionization equilibrium as well as nonequilibrium collisional ionization models do not explain the ion ratios at this temperature. The observed line strength ratios and line widths favor an ionization scenario in which both ion-electron collisions and UV photons contribute to the ionization in the gas. Such a model requires a low-metallicity of $\sim -1.7$~dex, ionization parameter of log~$U \sim -1.4$, a large total hydrogen column density of $N(\H) \sim 4 \times 10^{19}$~{\cmsq}, and a path length of $\sim 400$~kpc. The line of sight to Mrk 290 intercepts at the redshift of the absorber, a megaparsec scale filamentary structure extending over $\sim 20$ deg in the sky, with several luminous galaxies distributed within $\sim 1.5~h^{-1}$~Mpc projected distance from the absorber. The collisionally ionized gas phase of this absorber is most likely tracing a shock-heated gaseous structure, consistent with a few different scenarios for the origin including an over-dense region of the WHIM in the galaxy filament or highly ionized gas in the extended halo of one of the galaxies in the filament. In general, BLAs with metals provide an efficient means to study $T \sim 10^5 - 10^6$~K gas in galaxy halos and in the intergalactic medium. A substantial fraction of the baryons {\it missing} from the present universe is predicted to be in such environments in the form of highly ionized plasma. 

\end{abstract}

\section{Introduction}

Cosmic baryon census estimates have shown that in our present universe only $\sim 6 - 10$~\% of the primordial baryons are entrapped in galaxies. The bulk of the baryons have not collapsed into luminous structures but are in circumgalactic regions, and in the unvirialized large scale intergalactic filaments \citep{mulchaey96, fukugita98, fukugita04}. A significant fraction ($\sim 50$\%) of these baryons outside of galaxies has already been discovered by STIS/$HST$ and $FUSE$ UV spectroscopic observations of the $z < 0.5$ intergalactic medium (IGM). The low-$z$ IGM exists as a mix of multiple gas phases. The photoionized diffuse IGM with $T \lesssim 10^4$~K traced by the {\Lya} forest accounts for $\sim 30$\% of the baryonic mass density \citep{shull96, penton04, lehner06, danforth08}. The population of highly ionized metal line absorbers, notably {\OVI}  systems discovered closer to galaxies, possibly contain an additional $\sim 10$\% of the mass-fraction if their ionization is dominated by collisions of electrons with ions in the plasma \citep{tripp00b, richter04, danforth05, savage05, stocke06, danforth08, tripp08, narayanan10}. The remaining $\sim 60$\% of the baryonic mass-fraction in our present universe remains to be discovered. Hydrodynamic simulations of structure formation predict that this missing portion from the baryon inventory is in a gravitationally shock-heated phase with temperatures in the range of $T \sim 10^5 - 10^7$~K and densities of $n_{\H} \sim (0.1 - 10) \times 10^{-5}$~{\cc} \citep{cen99, dave01, valageas02}. The simulations convey that the reservoir of cooler intergalactic gas gets heated to warm - hot temperatures as a result of shocks produced during the gravitationally collapse of matter to form galaxies and clusters in the over-dense regions of the cosmic web. Discovering this warm-hot intergalactic medium (WHIM) is one of the main science drivers for the {\it Cosmic Origins Spectrograph} (COS) on $HST$ \citep{shull09}.

Currently, UV absorption line spectroscopy is the most promising observational technique for detecting the component of the WHIM with $T \sim 10^5 - 10^6$~K. Among metal lines, {\OVIdblt}~{\AA} are the most sensitive for observations of high ionization gas, due to the high cosmic abundance of oxygen and the large values of oscillator strengths of the $1032$~{\AA} and $1038$~{\AA} resonant lines. Under collisional ionization equilibrium conditions, {\OVI} ionization fraction reaches its peak value of $f_{\OVI} = N(\OVI)/N(\mathrm{O}) = 0.187$ at $T = 3.2 \times 10^5$~K \citep{gnat07}. Thus, {\OVI} can be a tracer of the {\it warm} phase (i.e. the lower temperature phase) of the shock-heated gas \citep{tripp01, savage02, danforth08}. Theoretical estimations predict  $\sim 20 - 30$\% of the missing baryonic matter is traced by {\OVI} absorption systems with $W_r(1032) \geq 20$~m{\AA} \citep{cen01}. However, observationally confirming the value of $\Omega_b(\OVI)$ has been a challenge. Detectable amounts of {\OVI} can also be produced in low-density gas ($n_{\H} \lesssim 10^{-5}$~{\cc}) photoionized and heated by a hard radiation field, in which case the absorber may not be tracing a large-scale structure \citep{savage98, prochaska04, thom08, tripp08, oppenheimer09}. The current sample of known {\OVI} systems seems to suggest that they are a heterogeneous mix, with evidence for photons dominating the ionization in some cases, and electron collisions in others. Detailed ionization modeling with tight constraints on column densities, Doppler widths and velocities of {\HI},  {\OVI} and other metal ions are necessary to understand the ionization mechanism in each absorption system and its physical origin \citep{savage05, narayanan10}. 

An alternative to using highly ionized metals is to search for absorption from the trace amounts of atomic hydrogen (ionization fraction, $f_{\HI} = N(\HI)/N(\H) < 10^{-5}$) present in the warm - hot gas. At $T \geq 10^5$~K, an absorber with a total hydrogen column density of $N(\H) \sim 10^{19}$~{\cmsq} will produce {\Lya} absorption with column densities of $N(\HI) \lesssim 10^{14}$~{\cmsq} and Doppler widths of $b(\HI) \gtrsim 40$~{\kms}. Accurate measurements of such broad lines in the quasar spectrum require high $S/N$, well-determined continua, and a good understanding of the fixed-pattern noise. Broad {\Lya} absorbers (BLAs) in the $z < 0.5$ universe were discovered in the existing archive of $HST$/STIS quasar spectra \citep{sembach04, richter06a, lehner07, danforth10}. The most serious systematic uncertainty associated with these BLA measurements is whether the observed broad line width of {\Lya} is dominantly thermal. Due to the shallow and broad shape of the absorption profile, particularly when the {\HI} column is low, it is often difficult to rule out, from moderate $S/N$ spectra, the possibility of line blending or turbulence contributing to line width. High $S/N$ spectroscopy is therefore crucial to confirm the thermally broadened Gaussian absorption profile characteristic of BLAs. At low-$z$ BLAs account for a baryonic mass-density fraction similar to galaxies or possibly even greater \citep{richter06a, danforth10}. Thus discovering more BLAs is central to the continuing search for the reservoirs of missing baryons in the present universe. Observations of the low-$z$ universe also opens up the possibility to search for an answer to the allied issue of how these baryons are distributed in relation to galaxies in large-scale structures. 

In this paper, we report on a BLA in the very nearby universe detected in the high $S/N$ spectra of Mrk~290 obtained with the $HST$/{\it Cosmic Origins Spectrograph}. The BLA is at the same velocity as an {\OVI} absorber identified in the $FUSE$ spectrum of the same sight-line. The BLA - {\OVI} system is tracing collisionally ionized gas associated with a large-scale galaxy filament. We describe in detail the COS observations (sec 2.1), the properties of the absorber (sec 3), and the details on the galaxies in the foreground field (sec 4). The physical properties of the absorber are investigated using the predictions made by different ionization scenarios (sec 5). Based on the observed properties and the ionization analysis, we speculate on the origin of the absorption system and its baryonic content (sec 6). The cosmology that we adopt is $\Lambda$CDM with $\Omega_\mathrm{m} = 0.3$, $\Omega_\Lambda = 0.7$ and a dimensionless Hubble parameter of $h = H_0/(100~{\kms}~\mathrm{Mpc}^{-1}) = 0.71$. 

\section{Observations}
\subsection{COS Observations}

The design capabilities of $HST$/COS are described in detail by \citet{green01, froning09} and in the updated COS Instrument Handbook \citep{cos2010}. The inflight performance of COS is discussed by \citet{osterman10} and in the numerous instrument science reports on the STScI COS website\footnote{http://www.stsci.edu/hst/cos/documents/isrs}.The COS observations of Mrk~290 were carried out on 28 October 2009 as part of cycle 17 COS GTO program ID 11541 (P.I. James Green).  The details of the observations are listed in Table 1. The G130M and G160M COS gratings were chosen to obtain spectra in the far ultraviolet bands. The G130M observations were split into four exposures of $\sim 964$~second integrations at grating central wavelength positions of $1291$~{\AA}, $1300$~{\AA}, $1309$~{\AA} and $1318$~{\AA}. Similarly, the G160M observations were composed of four $\sim 1200$~second integrations with central wavelength settings of $1589$~{\AA}, $1600$~{\AA}, $1611$~{\AA}, and $1623$~{\AA}. All exposures were obtained in the time-tagged, photon-counting mode. The different grating central wavelength configurations resulted in photons of the same wavelength getting recorded at different regions of the detector which subsequently reduced the effect of fixed pattern noise in the final coadded spectrum. This set up also helped to cover the $\sim 18$~{\AA} wavelength gap between segments A and B of the FUV detector for the various central wavelength settings.  

The reduction of the raw frames into 1D spectra was carried out using the COS pipeline software Calcos (v2.11) with the appropriate calibration reference files. The one-dimensional science data files generated by the pipeline had a dispersion scale of 0.01~{\AA} corresponding to 6 wavelength bins per resolution element. The individual integrations were coadded weighted by their respective exposure times. The final spectrum had a wavelength coverage from 1136~{\AA} to 1796~{\AA}. The signal-to-noise ratios (per 18~{\kms} resolution element) of this combined final spectrum at 1250~{\AA}, 1450~{\AA}, and 1650~{\AA} are 35, 24, and 16 respectively. 

Ghavamian {\etal}(2009) have determined the resolving power of the spectrograph by detailed modeling of the line spread function (LSF) at various wavelengths. The spectral resolution is found to be wavelength dependent with values in the range $\lambda/\Delta\lambda \sim 16,000 - 21,000$ for the G130M and G160M gratings, where $\Delta\lambda$ refers to the width at half strength of the LSF which has broad wings containing 25\% of the LSF area. The resolution is maximum at near-UV wavelengths and declines monotonically towards lower wavelengths. 

To determine zero-point offset in the COS spectrum we compare the velocity centroids of the $v_\mathrm{LSR} = -138$~{\kms} ISM absorption lines of {\SII}~$\lambda~1254$~{\AA}, {\FeII}~$\lambda 1608$~{\AA} and {\AlII}~$\lambda 1671$~{\AA} from Complex $C$ \citep{wakker06}.  These lines appear at $v_\mathrm{LSR} =  -159, -154$, and $-153$~{\kms} respectively in the COS spectrum. A velocity correction of $+ 17$~{\kms} was therefore applied to the COS data, which is approximately one resolution element. We estimate residual errors of $\sim 5$~{\kms} ($1~\sigma$) in the velocity calibration. 

\subsection{FUSE Observations}

The properties of the $FUSE$ satellite and its performance are described in detail by \citet{moos00} and \citet{sahnow00}. The processing of the $FUSE$ data for this sight line is described in great detail elsewhere \citep{wakker03, wakker06}. We only provide a short overview here. The archived $FUSE$ observations for Mrk~290 were obtained over the time period from 2000 to 2007 (exposure IDs: P1072901, D0760101, D0760102, E0840101, E0840101, E0840102).   The reduction of these data sets was carried out using version 2.4 of the $FUSE$ calibration pipeline. The shifts in wavelength and central velocities in the individual reduced exposures were corrected by aligning the Milky Way ISM lines with the LSR velocity of the 21 cm {\HI} absorption from Complex~$C$ ($v_\mathrm{LSR} = -138$~{\kms}). Once the velocity shifts were applied to each observation, the data in the LiF1A and LiF2B channels of the detector were combined together to produce the final spectrum. 

\section{Properties of the Absorber}

At a heliocentric velocity of $v_\mathrm{HELIO} = 3081$~{\kms}, the $z(\OVI) = 0.01027$ absorber is one of the nearest extragalactic {\OVI} absorption systems. It was discovered by Wakker \& Savage (2009) in the $FUSE$ spectrum of Mrk~$290$ through the identification of {\OVI}~$\lambda 1032$~{\AA} line. The detection was claimed as {\it uncertain} since the red member of the {\OVI} doublet is confused with the Galactic {\ArI}~$\lambda 1048$~{\AA} line. A hint of the {\OVI}~$\lambda 1038$~{\AA} came from the extra absorption wing seen at positive velocities of the Galactic {\ArI} line. The {\Lya} for this absorber was outside the $FUSE$ spectral window, and the {\Lyb} at this velocity suffers from Galactic {\CII} contamination. Nonetheless, \citet{wakker09} reported the $> 3~\sigma$ absorption detected at $\lambda \sim 1042.5$~{\AA} as {\OVI}~$\lambda 1032$~{\AA} after ruling out the possibility of the identified line being interstellar absorption or higher order Lyman lines along this sight line. The detection of {\Lya} in the COS G130M spectrum at $z(\Lya) = 0.01028$ confirms the \citet{wakker09} identification of this absorption system. 

In Figures 1a - 1b we show the continuum normalized profiles of {\Lya} and {\OVI} on a velocity scale centered at $z(\OVI) = 0.01027$. Also displayed are regions of the spectrum where absorption from other prominent ionic species are expected. The {\Lya} profile is spread over a velocity interval of $\Delta v = 400$~{\kms} in the rest frame of the absorber. The high $S/N$ COS data show two distinct components in {\HI}, separated in velocity by $v \sim 115$~{\kms}. Using the \citet{fitzpatrick97} software, we fit Voigt profiles to the {\Lya} feature. The fit convolves the model profile with the G130M instrumental spread function at the observed wavelength of the {\Lya} feature. It is important that the specific line spread function determined for COS be used to fit the profile in order to minimize the impact of the non-gaussian wings on observed line profiles \citep[see][for a detailed discussion]{coslsf}. 

The fit that best reproduces the observed absorption profile is a two component model with log~$[N(\HI)~{\cmsq}] = 14.35~{\pm}~0.01$~dex and $b(\HI) = 55~{\pm}~1$~{\kms} at $v = 2~{\pm}~5$~{\kms} and log~$[N(\HI)~{\cmsq}] = 14.07~{\pm}~0.01$~dex and $b(\HI) = 33~{\pm}~1$~{\kms} at $v = 116~{\pm}~5$~{\kms}. This model fit is displayed in Figure 2. The fit reveals some weak residual absorption in the positive and negative velocity wings of the {\Lya}. A broader, shallower {\HI} component could explain this excess absorption in the wings. However, because of blending with the stronger {\HI} component at $v = 2$~{\kms}, the fitting routine is not able to yield meaningful parameters for an extra component at $v \sim -125$~{\kms}. The contribution to the total column density from such a shallow component is going to be marginal ($W_r \sim 30$~m{\AA}, log~$[N(\HI)~{\cmsq}] \lesssim 13.0$) when compared with the contributions from the two stronger components at $v \sim 2$~{\kms} and $v \sim 116$~{\kms} respectively. 

The $b = 55$~{\kms} Doppler width for the $v = 2$~{\kms} {\HI} component suggests that this is a broad {\Lya} absorber (from here on, we often refer to this component in {\HI} as the BLA). If the line is purely thermally broadened, then $b = 55$~{\kms} line width implies a gas temperature of T = $1.8 \times 10^5$~K. The most significant systematic uncertainty that could affect a $b$-measurement is blending of closely separated kinematic components. The residual absorption in the blue wing of the BLA suggests some kinematic complexity. However, the BLA and the $v \sim 116$~{\kms} components of the {\Lya} profile appear highly symmetric suggesting very little influence of this low column density component on the measured $b$ values. 

We take note of the fact that the {\Lya} profile could be subjected to saturation that is unresolved by COS (FWHM~$\sim 17$~{\kms}). The presence of instrumentally unresolved saturation becomes evident when apparent column density profiles of different electronic transitions from the same energy level in an ion are compared \citep{savage91}. Such an analysis is  not feasible in the case of the $z = 0.01027$ absorber as the {\Lyb} covered by $FUSE$ is strongly affected by contamination from Galactic ISM {\CII} absorption. The {\Lyg}, which is $\sim 18$ times weaker compared to {\Lya} in line strength ($f\lambda$), falls at the edge ($\lambda \sim 982.5$~{\AA}) of the $FUSE$ coverage where the quality of the data is poor ($S/N \sim 4 - 6$ per resolution element). Even so, the {\HI} upper limit on column density of $10^{15.1}$~{\cmsq} given by the low $S/N$ {\Lyg} data suggests that there is no significant saturation in either component of {\Lya}. 

The COS spectrum offers coverage of several important low and high ions including {\CII}, {\SiII}, {\FeII}, {\SiIII}, {\CIV}, {\SiIV} and  {\NV} (see Figures 1a - 1b). The expected wavelengths for all these ions show non-detections at the 3~$\sigma$ significance level. We therefore quote in Table 2 upper limits on the column density for these nondetections. The lack of any low ionization absorption is indicative of high ionization conditions in the absorber. The  {\CIII}~$\lambda 977$~{\AA} line is covered by $FUSE$, but falls at the edge of the spectrum where the $S/N$ is poor (see Figure 1a). The line is not detected at $> 3~\sigma$ significance. 

A Voigt profile fit to the {\OVI}~$\lambda 1032$~{\AA} line yields a single component at $v = 1~{\pm}~5$~{\kms} with log~$[N(\OVI)~{\cmsq}] = 13.80~{\pm}~0.05$~dex and $b(\OVI) = 29~{\pm}~3$~{\kms}. We use a Gaussian of FWHM = 25~{\kms} in this fit model to account for the $FUSE$ instrumental smearing. The profile fit is displayed in figure 2. Using these fit parameters, we synthesize the {\OVI}~$\lambda 1038$~{\AA} line and superimpose the profile on top of the data in figure 2. The comparison shows the uncontaminated positive velocity segment of the {\OVI}~$\lambda 1038$~{\AA} absorption consistent with the {\OVI}~$\lambda 1032$~{\AA} profile. We also use the apparent optical depth (AOD) technique of \citet{sembach92}, to derive log~$[N_a(\OVI)~{\cmsq}] = 13.82~{\pm}~0.08$~dex and $b_a(\OVI) = 35~{\pm}~8$~{\kms} for the {\OVI}~$\lambda 1032$~{\AA} line. Our AOD measurements are similar to the Wakker \& Savage (2009) measurement of the same feature (see Table 2). The $b_a$ is the second moment of the apparent column density distribution, which when corrected for the $FUSE$ instrumental spread gives a $b$-value that is consistent with the profile fit measurement. 

In Figure 3, we plot the apparent column density profiles of {\Lya} and {\OVI}. It is evident that the {\OVI} absorption is kinematically well aligned with the BLA (i.e., the $v = 2$~{\kms} {\HI} component). The formal velocity difference between the line centers is $\Delta v = v(\OVI) - v(\Lya) = -1~{\pm}~5$~{\kms}. If the two ions are tracing the same gas phase, then the Doppler widths of $b(\HI) = 55~{\pm}~1$~{\kms} and $b(\OVI) = 29~{\pm}~1$~{\kms} implies a temperature of $T \sim 1.4 \times 10^5$~K with a non-thermal line broadening component of $b_\mathrm{nt} = 26$~{\kms}~\footnote{ We solve for the temperature assuming that the observed $b$ is a quadrature sum of the thermal and non-thermal $b$ values, i.e., $b_\mathrm{obs} = \sqrt{b_\mathrm{t}^2~+~b_\mathrm{nt}^2}$.}. The derived temperature suggests that the $v = 2$~{\kms} {\HI} component is by definition still a BLA (thermal $b > 40$~{\kms}), even though $\sim 50$\% of the observed line broadening is non-thermal. 

The wavelength region of the {\OVI}~$\lambda 1032$ corresponding to the narrower $v \sim 116$~{\kms} {\HI} component is contaminated by a feature that is unrelated to this system. The wavelength region corresponding to the {\OVI}~$\lambda 1038$ line shows no {\OVI} absorption associated with this {\HI} component. The $b(\HI) = 33~{\pm}~1$~{\kms} implies an upper limit on temperature of $T \leq 6.6 \times 10^4$~K, which is consistent with photoionization. No other low or high ionization metal lines are detected at the velocity of this {\HI} component. In Table 2, we separately list rest-frame equivalent width and column density upper limits for the various ions corresponding to the $v \sim 116$~{\kms} {\HI} absorption. 

\section{Absorber - Galaxy Association}

A search through the NED database reveals several extended sources within $1000$~{\kms} of the BLA - {\OVI} absorber in the foreground field of Mrk 290. The {\it left panel} of Figure 4a shows the projected distribution of these galaxies within a $\sim 16~h^{-1}$~Mpc region. The distribution pattern suggests that the sight line is intercepting at the velocity $v_\mathrm{HELIO} = 3081$~{\kms} of the absorber an over-dense filamentary structure in the cosmic web, with the galaxies organized around the axes of the filament. The megaparsec scale filamentary structure extends $\sim 15^{\circ}$ on the sky. In Figure 4a, we also plot the position of other AGN sight lines in the field. Among those, Mrk 876 lies $9.7^{\circ}$ ($\sim 7~h^{-1}$~Mpc) further along the filament and shows {\Lya}, {\Lyb} and {\OVI} absorption at $v_\mathrm{HELIO} = 3480$~{\kms} most likely associated with the filament \citep{wakker09}. The PG~$1626+554$ sight line also probes this filament at a separation of $7^{\circ}.5$ ($\sim 5.6~h^{-1}$~Mpc) in the sky in the opposite direction to Mrk 876, but does not show any {\HI} absorption near the expected velocity. The other AGNs shown in the Mrk 290 field (e.g. Mrk~279, Mrk~817, PG~$1351+640$) lie farther away from the filament and do not have any metal line or {\HI} absorption detected near $v_\mathrm{HELIO} \sim 3100$~{\kms}.  

The Mrk 290 field is inside the SDSS footprint. The galaxy sample in this region of this sky is nearly 100\% complete down to a magnitude of $g = 17.5$ which corresponds to $\sim 0.03L^*$ \citep{blanton03}. From the more than two dozen galaxies within a projected distance of $\sim 1.5 h^{-1}$~Mpc from the absorber, NGC~$5981$, NGC~$5982$, NGC~$5985$, NGC~$5987$ and NGC~$5989$ are members of the loose group GH~158 \citep{geller83}. The galaxy distribution shows at least nine other galaxies within $\sim 1.5~h^{-1}$~Mpc impact parameter and $\Delta v = 300$~{\kms} systemic velocity with respect to the absorber, but smaller in size and brightness compared to the GH~158 group members. Some of the other brighter, larger galaxies along this filament are also members of  other galaxy groups (e.g. GH151). 

From all these known extended foreground sources, the luminous galaxy NGC~$5987$ is closest in impact parameter to the absorber. NGC~5987 [$\alpha (\mathrm{J} 2000) = 15^h39^m57.4^s, \delta (\mathrm{J} 2000) =  +58^{\circ}04^{\prime}46^{\prime\prime}$] is at a projected distance of $\rho = 424~h^{-1}$~kpc from the sight line. The galaxy has a redshift of $z = 0.0100$ \citep[v$_\textrm{\scriptsize{HELIO}} = 3010$~{\kms}][]{devaucouleurs91} which gives it a $|\Delta v| = 71$~{\kms} systemic velocity with respect to the BLA - {\OVI} absorber. The galaxy has an estimated isophotal major diameter of $D_{25} = 51.7$~kpc, and is one of the largest spiral galaxies in the nearby universe \citep{romanishin83}.  Although the impact parameter is large, the small velocity separation and the large size of the galaxy makes it possible for the absorber to be in some way related to NGC~$5987$'s circumgalactic environment. Such an inference is consistent with statistical results from a number of past absorber-galaxy surveys where $> 0.25L^*$ galaxies were found within $\sim 500$~kpc of {\OVI} absorbers \citep{tripp00a,tripp00b,tumlinson05,stocke06,cooksey08,chen09,lehner09,wakker09} and also with results from cosmological simulations of galaxies and {\OVI} absorbers \citep{ganguly08}.

NGC 5987 is a spiral galaxy of Sb morphology with a $B$-band magnitude of $B = 12.72$ or $M_B - 5$~log $h = -20.4$ \citep{devaucouleurs91}. This corresponds to a luminosity of $L \sim 2~L^*$ assuming $M^*_B = -19.6$. No published spectrum is available for this galaxy. In Figure 4b, we display the orientation of the galaxy with respect to the quasar sight line. The large impact parameter of the absorber with respect to the galaxy disk results in low surface flux of UV photons from the galaxy incident on the absorber. Thus, we do not expect the galaxy's radiation field to significantly influence the ionization of hydrogen or metals in the absorber \citep{fox05,narayanan10}. NGC 5987 also does not appear in the ROSAT all-sky survey bright source catalogue \citep{voges99} and therefore is unlikely to harbor an active AGN that could photoionize the surrounding intergalactic gas. 

Given the significant excess of galaxies with $v_\mathrm{HELIO} \sim 3080$~{\kms} in the field around Mrk~290, and the pencil beam size region probed by the BLA - {\OVI} system ($v_\mathrm{HELIO} = 3081$~{\kms}), it will be difficult to conclude whether the absorption is tracing the gaseous halo of one or the other galaxy in the filament, or an intergalactic gas cloud that is part of the large-scale filament. We return to this discussion in Sec 5. 

\section{Ionization Conditions in the Absorber}

Understanding the ionization, physical conditions and metalicity in the absorber is crucial to discriminate between the different possible explanations for its astrophysical origin.  We explore a range of parameter space in both photoionization and collisional ionization models and assess whether the models are physically realistic. We focus our modeling on the $v \sim 2$~{\kms} BLA component of the {\HI} absorption which has {\OVI} detected at the same velocity. No metals are detected for the other {\HI} component at $v \sim 116$~{\kms} and therefore useful constraints on the physical parameters cannot be obtained for this component from ionization modeling. 

\subsection{Photoionization}

The $T = 1.4 \times 10^5$~K temperature implied by the different BLA and {\OVI} line widths is much higher than the kinetic temperature expected for a purely photoionized gas. Nonetheless, to elaborate on the difficulties associated with photoionization in explaining the origin of these lines, we investigate some best-fit photoionization models and their predictions. We use the photoionization code Cloudy [ver.C08.00, \citet{ferland98}] to solve for equilibrium models that reproduce $N(\OVI)$ and the limits on all the other ions. The source of ionization is assumed to be dominated by the extragalactic background radiation at $z = 0.01027$ whose shape and intensity is as modeled by \citep{haardt01}. The UV background spectrum used has contributions from quasars and from the population of high-$z$ young star forming galaxies. The photoionization models are calculated for different ionization parameters log~$U$\footnote{Ionization parameter is defined as the ratio of the number density of photons with E $ \geq 13.6$~eV to the hydrogen density, $U = n_{\gamma}/n_{\H}$},  metallicity and the observed $N(\HI) = 10^{14.35}$~{\cmsq} in the broad {\HI} component at $v = 1$~{\kms}. The $N(\OVI) = 10^{13.80}$~{\cmsq} and the limiting column density ratios of $N(\CIV)/N(\OVI)$, $N(\NV)/N(\OVI)$, $N(\SiIV)/N(\OVI)$ and $N(\CIII)/N(\OVI)$ serve as additional constraints for the models. In the models, we assume the revised solar elemental abundance ratios of [C/H]$_\odot$ = -3.57~dex, [O/H]$_\odot$ = -3.31~dex, [N/H]$_\odot$ = -4.17~dex and [Si/H]$_\odot$ = -4.49~dex given by Asplund {\etal}(2009) 

Shown in Figure 5 are the photoionization model predictions for column densities of the various ionic species, as a function of the ionization parameter. The observed $N(\CIV)/N(\OVI)$ and $N(\NV)/N(\OVI)$ limiting ratios restrict the ionization parameter to log~$U \gtrsim -0.6$, corresponding to $n_{\H} \lesssim 6 \times 10^{-6}$~{\cc} and metallicity to -1.8 dex of solar. The $N(\HI) = 10^{14.35}$~{\cmsq} and the log~$U$ dependent ionization fraction of $f_{\HI} \lesssim 10^{-5}$ implies an absorbing region  with a total baryonic column density of $N(\H) \geq 4 \times 10^{19}$~{\cmsq}. The very low density and high total hydrogen column density results in a line of sight thickness of $L \gtrsim 3$~Mpc for the absorbing structure. If the absorber is unvirialized, the large path length would result in lines with $v > H(z)~L \sim 210$~{\kms} width due to Hubble flow, which is inconsistent with the observed {\HI} and {\OVI} velocity widths. If the absorbing structure is well within the virialized halo of NGC~$5987$, it would have decoupled from cosmic expansion. The kinematic properties of the BLA would then largely depend only on the internal velocity field along the line of sight and the rotational velocity of the galaxy halo. It would be remarkable for a megaparsec scale absorbing region to not have any velocity field capable of inducing asymmetry or kinematic complexity in the line profiles. To the contrary, we find no such evidence in the observed line profiles of the BLA and {\OVI} which appear kinematically simple, symmetric and Gaussian. The predictions of the models are further complicated by the fact that the photoionization equilibrium temperature of $\sim 0.6 \times 10^5$~K is significantly smaller than the $T = 1.4 \times 10^5$~K given by the different BLA and {\OVI} line widths. Thus photoionization does not appear as a dominant process controlling the ionization state. We therefore turn to the alternate scenario of collisional ionization. 

\subsection{Collisional Ionization}

\subsubsection{Collision Ionization Equilibrium Models}

The temperature suggested by the observed line widths is consistent with collisional ionization conditions. First we consider the simplest scenario of collisional ionization equilibrium (CIE) in which at a constant kinetic temperature the fraction of ions remain time-independent. In Figure 6, we compare observations with column density predictions from CIE models of \citep{gnat07} for a metallicity of 1/10 solar. We find that CIE models fail to explain all observed high ion column density limits simultaneously for acceptable temperatures. The predicted $N(\OVI)$ is greater than $N(\CIV)$ and $N(\NV)$ for $T \gtrsim 2 \times 10^5$~K, which is higher than the temperature given by the line widths of the BLA and {\OVI}. Between the temperature range of $T = 3.2 \times 10^5$~K, where the {\OVI} ionization fraction peaks and $T = 10^5$~K, the {\OVI} fraction declines rapidly by a factor of $\sim 4$~dex. Thus, at $T = 1.4 \times 10^5$~K, 0.1Z$_\odot$ metallicity and $N(\HI) = 10^{14.35}$~{\cmsq}, the CIE predicted $N(\OVI)$ is a factor of $\sim 3$~dex smaller than the measured value and the $N(\CIV)$ and $N(\NV)$ are significantly overproduced for solar relative elemental abundances. 

\subsubsection{{\it Hybrid} of CIE \& Photoionization Models}

Even when conditions are favorable for collisional processes to dominate the ionization state, a realistic modeling of absorption systems should take into account the effect of UV metagalactic flux from quasars and star forming galaxies on the absorber \citep{danforth06,richter06b, tripp08,danforth10}. We therefore compute a series of hybrid models using Cloudy, allowing both collisional and photoionization reactions to contribute to ion fractions. The temperature in Cloudy is set to $T = 1.4 \times 10^5$~K given by the different BLA - {\OVI} line widths. The model that Cloudy would converge on will have the ionization and recombination in equilibrium even though the heating and cooling rates are out of balance, due to the fixed temperature. In Figure 7 we show the column density predictions for these hybrid models at $T = 1.4 \times 10^5$~K and  different log~$U$. The density of the absorber in these models is determined by the input number density of hydrogen ionizing photons and the ionization parameter. The hybrid models that reproduce the measured $N(\OVI)$ and the column density ratios of {\OVI} with {\CIV} and {\NV} require [Z/H] $\sim -1.9$~dex, log~$U \sim -0.6$, $n_{\H} \sim 4 \times 10^{-6}$~{\cc}, $N(\H) \sim 1.8 \times 10^{20}$~{\cmsq} and thus absorber path lengths $L \sim 12$~Mpc. The expected line width from Hubble broadening for such a large path length is incompatible with the observed {\HI} and {\OVI} line widths and therefore a CIE - photoionization hybrid model can be rejected. 

\subsubsection{Non-equilibirum Collisional Ionization Models}

The temperature $T \sim 10^5$~K implied by the data is in the regime where radiative cooling is efficient. The presence of metals can enhance the rate of cooling of gas such that significant departures of ion fractions from CIE will happen at these temperatures. The time dependent ionization fractions in a gas radiatively cooling from an equilibrium temperature of $T \sim 10^6$~K were calculated by Gnat \& Stenberg (2007). In Figure 8, we illustrate the non-equilibrium collisional ionization (noneq-CI) curves for such a cooling gas flow using their model for metallicites of Z$_\odot$ and 0.1Z$_\odot$. In non-equilibrium cooling models, the recombination process lags behind the declining temperature such that the ionization fraction for high ions will be higher compared to CIE for temperatures in the range $10^{4.5} - 10^{5.3}$~K. The effect is pronounced for higher gas metallicities. At solar metallicity and $T = 1.4 \times 10^5$~K, the {\OVI} ion fraction in noneq-CI models is $f(\OVI) = 2.63 \times 10^{-3}$, a factor of 1.4~dex larger than the {\OVI} ion fraction at CIE. From Figure 8, it can be inferred that for $Z \sim 0.3~Z_\odot$, the $N(\OVI)$ given by the model will be in agreement with the observed value for $T = 1.4 \times 10^5$~K. However, for the same model, the predicted $N(\CIV) \sim N(\OVI)$ whereas observations suggest $N(\OVI)/N(\CIV) \gtrsim 8$. The {\CIV} prediction from this model would be consistent with the data only if the carbon to oxygen abundance ratio is reduced from solar by $\sim 1$~dex. Thus, a simple noneq-CI scenario may not explain in a satisfactory manner the ionization in this absorber. However by including in these noneq-CI models the influence of UV photons from the extragalactic background radiation, an acceptable ionization solution can be obtained. This is elaborated in the following subsection.

\subsubsection{Hybrid of noneq-CI \& Photoionization Models}

In Figure 9, we display the changes to {\OVI} and {\CIV} fractions when collisions of ions with both electrons and photons are simultaneously influencing the ionization in the gas. The models in Figure 9 were computed using Cloudy for a fixed temperature of $T = 1.4 \times 10^5$~K given by the BLA - {\OVI} line widths and a fixed intensity of the \citet{haardt01} (QSOs and galaxies) extragalactic radiation field determined for the redshift of the absorber. The models are therefore a hybrid of equilibrium photoionization and equilibrium collisional ionization reactions. The changes perceived for the ion ratios in these hybrid models can be used to estimate how the ion ratios would be affected in a noneq-CI cooling gas scenario when photoionization is included. It should not be forgotten that several broad assumptions are inherent in these ionization models (plane-parallel slabs of uniformly cooling gas with no internal temperature gradient). Few real absorbers would resemble closely with these conditions. We should only expect these computational models to converge on a general acceptable solution rather that an exact fit to the full range of observed ionic column densities. Our correction allowing for the effects of photoionization on the noneq-CI cooling gas is only approximate. A proper calculation would need to follow the changing ionizing conditions in gas cooling from $T \sim 10^6$~K to $T \sim 1.4 \times 10^5$~K when simultaneously subjected to the extragalactic radiation field. 

From the hybrid models shown in the {\it left panel} of Figure 9, we draw the following conclusions. For low values of log~$U$ (i.e. high density), ionization by photons does not contribute significantly to the {\OVI} to {\CIV} ratio. For log~$U < -4$, $N(\OVI)/N(\CIV)$ is similar to its CIE value of $N(\OVI)/N(\CIV) \sim -3.4$~dex at $T = 1.4 \times 10^5$~K. With an increase in ionization parameter (i.e., a large number of ionizing photons per electron), photoionization begins to increasingly influence the ionization fraction of {\OVI} such that over the interval $-4.5 \leq$~log~$U \leq -1.5$, the $N(\OVI)/N(\HI)$ and $N(\OVI)/N(\CIV)$ ratios are raised by a factor of $\sim 3$~dex. This region of the parameter space corresponds to the interval that separates models where {\OVI} is predominantly collisionaly ionized from models where photoionization begins to predominate. Within this log~$U$ interval the {\CIV} ionization fraction remains the same as CIE and begins to decline at log~$U > -2.5$. We find that the change to the $N(\OVI)/N(\CIV)$ ion column density ratios seen in Figure 9 ({\it left panel}) is not sensitive to metallicity. This increase in the ionization fraction of {\OVI} from pure collisional values, due to the introduction of the ionizing radiation field can be used to interpret hybrid models of noneq-CI and photoionization. 

For [Z/H] = -2.0, the noneq-CI model predictions at  $T = 1.4 \times 10^5$~K are $N(\OVI) \sim 10^{10.6}$~{\cmsq}, $N(\CIV) \sim 10^{13}$~{\cmsq}, $N(\NV) \sim 10^{12.3}$~{\cmsq} (refer {\it right panel} Figure 9). The {\NV} prediction is already consistent with the upper limit given by the data, whereas {\CIV} is overproduced by a small amount (0.1~dex higher), and {\OVI} underproduced by $\sim 3.2$~dex. In the hybrid model shown in the {\it left panel} of Figure 9, we can see that the {\OVI} ionization fraction is enhanced from noneq-CI by $\sim 3.2$~dex for log~$U \sim -1$, whereas the {\CIV} ionization fraction is diminished by $\sim 0.5$~dex, making both column density predictions consistent with the measurements. However, the density $n_{\H} \sim 10^{-5}$~{\cc}, and total hydrogen column density of $N(\H) \sim 6 \times 10^{19}$~{\cmsq} corresponding to this ionization parameter implies a line of sight of $\sim 2$~Mpc through the absorbing region. Such a large path length is inconsistent with the observed {\Lya} and {\OVI} line widths. Lowering the ionization parameter would decrease the amount of {\OVI} and increase the amount of {\CIV} making the predictions disagree with measurements.

For the {\it hybrid} model to agree with measurements, the log~$U$ has to be reduced (thereby narrowing the path length) simultaneously raising the metallicity in the collisionally ionized phase. For [Z/H] = -1.6, the $N(\OVI)$ will increase by $\sim 0.7$~dex from its [Z/H] = -2.0 value (see Figure 9) whereas the $N(\CIV)$ will increase by $\sim 0.4$~dex. In the presence of ionizing photons the predicted $N(\OVI)$ will increase by an additional $2.5$~dex making it equal to the measured value for log~$U \sim -1.5$. For this {\it hybrid} model, the $N(\CIV)$ will be $\sim 0.4$~dex smaller than the noneq-CI predicted value of $N(\CIV) \sim 10^{13.4}$~{\cmsq} making it approximately consistent with the $3~\sigma$ upper limit of  $10^{12.9}$~{\cmsq}. The log~$U \sim -1.5$ corresponds to $n_{\H} \sim 4 \times 10^{-5}$~{\cc} and a total hydrogen column density of $N(\H) \sim 4 \times 10^{19}$~{\cmsq} implying a path length of $\sim 320$~kpc for the absorbing region. Such a path length is consistent with the {\Lya} and {\OVI} line widths. The \citet{gnat07} noneq-CI models (shown in Figures 8 and 9) are available only for values of metallicity that differ by 1 dex. We have therefore used approximate scaling estimates for the noneq-CI column densities at [Z/H] = -1.6. 

There is a narrow range in metallicity and ionization parameter space where the {\it hybrid} model predictions will be consistent with the measured column densities and line widths. The values of [Z/H] = -1.6 and log~$U \sim -1.5$ are an upper limit on the metallicity and a lower limit on the ionization parameter, respectively. Increasing the metallicity would result in {\CIV} being overproduced from the {\OVI} phase. For a lower metallicity of [Z/H] = -1.8, the log~$U \sim -1.3$ in order to recover the observed $N(\OVI)$. This {\it hybrid} model would also be consistent with the $N(\CIV)$ upper limit. However, the suggested path length is $\sim 500$~kpc. Thus [Z/H] = -1.8 and log~$U \sim -1.3$ are therefore lower and upper limits respectively. Thus, within the framework of this simple noneq-CI and photoionization {\it hybrid} model, the BLA - {\OVI} gas phase has a metallicity of [Z/H] = $-1.7~{\pm}~0.1$~dex and ionization parameter log~$U = -1.4~{\pm}~0.1$~dex. The implied total hydrogen column density is very large ($N(\H) \sim 4 \times 10^{19}$~{\cmsq}) and therefore interesting. It is intriguing that even though the BLA is tracing collisionally ionized hydrogen at $T \sim 1.4 \times 10^5$~K, in the {\it hybrid} model much of the {\OVI} is produced through photoionization. 

It is important to emphasize here that the estimated metallicity and ionization parameter are dependent on the simplistic assumptions intrinsic to the ionization models and hence should not be considered as very robust. For example, the predicted $\sim 2$\% solar metallicity and $n_{\H} \sim 4 \times 10^{-5}$~{\cc} density would imply a radiative cooling time-scale of $\sim 25$~Gyr for the gas to cool from an initial temperature of $T \sim 10^6$~K \footnote{The radiative cooling time for a low-density plasma is given by $t_\mathrm{cool} = 1.5kT/(n_{\H}\Lambda(T))$ \citep{sutherland93}, where $k$ is the Boltzmann's constant and $\Lambda(T)$ is the temperature dependent radiative cooling efficiency of plasma. For gas at $T = 10^6$~K with $\sim 2$\% solar metallicity, $\Lambda \sim 10^{-24}$ erg s$^{-1}$ cm$^3$ \citep{gnat07}}. However, this time-scale estimation depends on the density and metallicity of the gas, and the initial temperature from which the gas begins cooling, all of which are not well determined for this absorber. A lower ionization parameter (higher density), higher metallicity or lower initial temperature would all result in lower cooling time-scales values that are consistent with the Hubble time-scale. 

The above {\it hybrid} model also has the limitation that it does not take into account the enhanced metal line cooling due to the increase in line emissions from excess {\OVI} produced by photoionization.  A $\sim 3$~dex increase in the {\OVI} column density could lower the cooling time scales thereby altering the ionization states of {\CIV} and {\OVI}. A more accurate treatment of the {\it hybrid} model should incorporate photoionization by external radiation into the \citet{gnat07} ionization calculations and subsequently follow the ion fractions as the gas cools from $T \sim 10^6$~K.  This is a computationally elaborate task and is beyond the scope of this paper.  

\section{Astrophysical Origin of the BLA - {\OVI} Absorber} 

In contemplating the nature of this BLA - {\OVI} absorber, we are guided by two key insights,  namely (1) the absorption is produced by gas at $T = 1.4 \times 10^5$~K whose ionization is likely governed by both collisional non-equilibrium and photoionization reactions (2) the absorbing structure is part of a large-scale filament of galaxies with several luminous galaxies within a $\sim 1.5~h^{-1}$~Mpc projected distance. In view of these, a few different scenarios are possible for the origin of the BLA - {\OVI} absorption. The information obtained from probing a pencil beam region of the absorber may not be sufficient to discriminate between these different physical scenarios. Given the uncertainties encountered in the ionization models, we treat the inferences from the models as secondary to what the data is directly suggesting.

Starburst driven outflows with large velocities from galaxies can shock-heat and increase the ionization of the surrounding gas to produce {\OVI} \citep{shull79, heckman01, indebetouw04, oppenheimer09}. Observationally such a physical scenario is supported by the detections of {\OVI} systems with [O/H] $\geq 0.2$~dex in the Galactic halo as well as the extended halos of other galaxies within $\sim 500$~kpc \citep{savage02, fox04, tumlinson05, stocke06, wakker09}. However, gas associated with chemical feedback from star-formation typically have metallicity higher than the $\sim -1.7$~dex given by the noneq-CI and photoionization {\it hybrid} models for the BLA - {\OVI} absorber. 

A shock-heated origin for {\OVI} was suggested by \citet{stocke06}. In their sample of {\OVI} absorbers and galaxies, $43$\% (16/37) of the absorbers were associated with pairs of {\Lya} clouds within $\Delta v = 50 -200$~{\kms} of each other. Both the {\OVI} and the {\Lya} absorptions [$N(\HI) \geq 10^{13.2}$~{\cmsq}] were also within $500$~kpc of $L^*$ and 100~kpc of 0.1~$L^*$ galaxies. About half of these {\OVI} in {\Lya} pairs were also found associated with galaxy groups. The velocity differences between the {\Lya} clouds were sufficient to create post-shock temperatures for {\OVI} to be produced. The $z = 0.1212$ {\OVI} absorber towards H~$1821+643$ is another example in which the hydrogen is seen in a cluster of {\Lya} absorbers. In this example, the various {\Lya} components were within $\sim 100$~{\kms} of each other \citep{tripp01}. The {\HI} component kinematically aligned with the {\OVI} in that absorber was a BLA with $b \sim 85$~{\kms} (see Table 3). Several galaxies with impact parameters ranging from 144 kpc to 3.1 Mpc were also identified within $|\Delta v| \sim 500$~{\kms} of the absorber implying that the sight line was intercepting a galaxy group \citep{tripp98, tripp00b}. Collisional ionization was inferred as the likely mechanism for the ionization in the {\OVI} - BLA gas phase in that system \citep{tripp01}. The $z = 0.01027$ {\OVI} absorber is also associated with a pair of {\Lya} absorbers separated by $|\Delta v| \sim 115$~{\kms}. Gas clouds moving at those relative velocities, colliding with one another, can shock-heat gas to temperatures of $T_s =  1.38 \times 10^5 (v_s/100)^2 \sim 1.8 \times 10^5$~K \citep[assuming $\gamma = 5/3$][]{draine93}, consistent with the temperature for the BLA - {\OVI} gas phase. Shocks can therefore be one of the mechanisms for the ionization in the absorber. 

In a study of {\OVI} absorbers and galaxies along select quasar fields, Chen \& Mulchaey (2009) find some {\OVI} absorbing galaxies with disturbed morphologies indicating past events of mergers or interactions. They argue that such galaxies likely have a higher cross-section for {\OVI} absorption due to tidally stripped gas contributing to the halo gas content. The BLA - {\OVI} absorber is also in a galaxy excess environment ($\sim 30$ galaxies within $1.5~h^{-1}$~Mpc projected distance) where the probability of galaxy pair interactions is high. Consequently, it is plausible that the absorption is from a shock-heated fragment of gas tidally stripped by NGC~5987 from one of the smaller galaxies in the large-scale filament.  The oxygen abundance in the ISM of low-mass irregular galaxies is found to be low \citep[$< 0.1Z_\odot$, ][]{skillman89a, kennicutt01}. The tidally stripped interstellar gas of such a dwarf satellite would be consistent with the low-metallicity for the BLA - {\OVI} absorber given by the {\it hybrid} models. The possible difficulty associated with this scenario is that tidally stripped gas is likely to produce absorption that is kinematically complex \citep{gibson00, wakker03, fox10}, which is not seen in the BLA and {\OVI} profiles.

The BLA - {\OVI} absorption can also be associated with the  warm-hot intergalactic gas in the galaxy filament. As the dominant phase of baryons at $z \sim 0$, the WHIM is expected to be the primary reservoir of gas feeding mass into galaxies. The $T \gtrsim 10^5$~K suggested by the data for the absorber corresponds to the temperature predicted for the moderately over-dense cooler component of the WHIM \citep{keres09}.  The BLA - {\OVI} absorber thus presents a unique opportunity to probe the interaction of WHIM tied-in with galaxy halos. The $\sim 2$\% solar metallicity for the absorber, given by the {\it hybrid} model, is consistent with the range measured for the low-$z$ IGM \citep[e.g.,][]{danforth05}. Even if the sight line to Mrk 290 is intercepting a nearby galaxy's (e.g., NGC 5987) extended halo, the absorption could still be from a shock-heated gas cloud accreted from the nearby WHIM. Such inflows of matter from WHIM are expected in the cold gas accretion models of galaxy formation \citep{keres05, brooks09, keres09}. The presence of paired {\Lya} absorbers, and an origin in a galaxy filament, makes the absorber an excellent candidate for shock-heated WHIM \citep{stocke06}.

\section{BLAs with Metals As A Tracer Of WHIM}

In this section we remark on why BLAs with metals are an important class of absorption system. Considerable observational effort has been invested to detect warm - hot collisionally ionized gas in the low-$z$ universe, as it is predicted to be the dominant reservoir phase of ordinary matter \citep[e.g.,][]{bregman07}. The broad-{\Lya} absorbers (by definition thermal $b(\HI) > 40$~{\kms}) offer a metallicity independent means to trace warm - hot gas \citep{richter06a, danforth10}. However the biggest uncertainty associated with the temperature measurement in BLAs is in ascertaining the fraction of line broadening that is thermal. Blending of closely separated absorption components and turbulence within the gas can significantly contribute to the broadening of the absorption. The extent of this contribution is not readily assessable from the broad {\HI} profile itself. Detecting metals associated with BLAs helps to resolve this problem. From the combined $b$-values for the metal line and {\HI}, one can solve for the thermal and non-thermal contributions to the line width and establish a tight constraint on the temperature of the gas. The temperature measurement is essential to determine with a certain level of precision the ionization fraction of {\HI} and the baryon content in BLAs. This value of $N(\H)$ can be used in conjunction with single-phase solution to also get an approximate measure of the abundance of metals in the gas. The absorption system discussed in this paper is an example in which such an analysis was possible. 

Metals have been detected in other BLAs. In table 3 we provide a partial list of BLAs with associated metal line detections with $|v (\mathrm{BLA}) - v (\mathrm{metal})| \leq 10$~{\kms}. The close alignment between the {\HI} and metal absorption (usually {\OVI}) can be evidence for the origin of both species in a single gas phase. For each example listed in table 3, the temperature derived from the $b(\HI)$ and $b(\OVI)$ values is $T \geq 10^5$~K, suggesting a dominant thermal broadening component for the line widths. The ionization in the gas phases producing the BLA - {\OVI} absorption in each of these cases could also possibly be understood using the nonequilibrium {\it hybrid} scenarios similar to the $z = 0.01027$ system described in this paper. 

\subsection{Summary}

We have presented HST/COS high $S/N$ observations of the $z(\OVI) = 0.01027$ broad {\Lya} - {\OVI} absorber in the spectrum of the Seyfert 1 galaxy Mrk~290. The absorber was discovered by Wakker \& Savage (2009) from the identification of the {\OVI}~$\lambda 1032$~{\AA} line in the $FUSE$ spectrum, which did not cover {\Lya}. The main conclusions from our analysis are as follows : 

(1) The COS high $S/N$ G130M spectra show {\Lya} absorption associated with the $z(\OVI) =0.01027$ system. In addition, the combined COS G130M and G160M spectra cover lines from {\CII}, {\SiII}, {\FeII}, {\CIII}, {\SiIII}, {\SiIV}, {\CIV} and {\NV}. All these ions are non-detections at the $3~\sigma$ significance level. The {\OVIdblt}~{\AA} lines are detected in the FUSE spectrum of Mrk~$290$. 

(2) The {\Lya} shows absorption from two distinct components with log~$[N(\HI)~{\cmsq}] = 14.35~{\pm}~0.01$, $b(\HI) = 55~{\pm}~1$~{\kms}, and log~$[N(\HI)~{\cmsq}] = 14.07~{\pm}~0.01$, $b(\HI) = 33~{\pm}~1$~{\kms} at velocities of $v = 2~{\pm}~5$~{\kms} and $v = 116~{\pm}~5$~{\kms} in the rest frame of the absorber. The $v = 2$~{\kms} component is by definition a BLA ($b > 40$~{\kms}). The BLA absorption profile is highly symmetric and Gaussian, with a dominant thermal broadening component of $b_\mathrm{thermal}/b_\mathrm{total} \sim 0.5$. 

(3) For the {\OVI}, we measure line parameters of log~$[N(\OVI)~{\cmsq}] = 13.80~{\pm}~0.05$, $b(\OVI) = 34~{\pm}~5$~{\kms}. The {\OVI}~$\lambda 1032$~{\AA} absorption is kinematically well aligned with the BLA, with a $\Delta v = v(\OVI) - v(\Lya) = -1~{\pm}~5$~{\kms}. The different $b$-values for the BLA and {\OVI} yields a kinetic temperature of $T = 1.4 \times 10^5$~K in the gas, implying that $\sim 50$\% of the line broadening is thermal.  

(4) The Mrk~290 sight line is intercepting a filament of galaxies at the location of the absorber. The megaparsec scale filamentary structure extends over $\sim 20$ deg in the sky, with several luminous galaxies distributed within $\sim 1.5~h^{-1}$~Mpc projected distance and $\Delta v = 1000$~{\kms} systemic velocity of the absorber. Among the galaxies, NGC~5987 is closest in impact parameter and velocity to the absorber. The galaxy is at a projected distance of $\rho = 424~h^{-1}$~kpc, at a systemic velocity of $|v| = 71$~{\kms} and oriented almost edge on with respect to the absorber. NGC 5987 is an Sb galaxy with an isophotal major diameter of $D_{25} = 51.7$~kpc and a B-band magnitude of B = 12.72 or $M_B - 5$ log $h = -20.4$ corresponding to a luminosity of $\sim 2L^*$. 

(5) Photoionization models do not provide physically realistic solutions for the ion ratios in the absorber. The model that best satisfies the line strength ratios requires the absorbing region to have a very low density of $n_{\H} \lesssim 6 \times 10^{-6}$~{\cc} and a path length of $ L \gtrsim 3$~Mpc. Such a large path length would result in Hubble flow broadening that is inconsistent with the observed velocity widths of {\HI}, {\OVI}, and their relatively simple and  symmetric line profiles. Also the temperature prediction of $T \sim 0.6 \times 10^5$~K made by the best-fit photoionization model is significantly smaller than $T = 1.4 \times 10^5$~K obtained from the different BLA and {\OVI} $b$-values.

(6) We find that the existing simple CIE models, noneq-CI cooling models as well as {\it hybrid} models of CIE and photoionization do not yield a satisfactory explanation for the ionization in the absorber. The predictions most consistent with the observations are  from {\it hybrid} models of non-equilibrium collisional ionization and photoionization. Within the frame work of such a {\it hybrid} model, a metallicity of $-1.7~{\pm}~0.1$~dex and ionization parameter of $-1.4~{\pm}~0.1$~dex corresponding to a density of $n_{\H} \sim 4 \times 10^{-5}$~{\cc}, and a baryonic column density of $N(\H) \sim 4 \times 10^{19}$~{\cmsq} fits the observed ionic ratios at $T = 1.4 \times 10^5$~K.  The value of the derived metallicity is highly dependent on the validity of the model. In this model, the BLA is tracing collisionally ionized hydrogen whereas much of the {\OVI} is produced via photoionization. 

(7) The galaxy filament and the detection of low-metallicity gas at $T = 1.4 \times 10^5$~K suggest that this absorber is tracing a shock-heated gaseous structure, consistent with a few different scenarios including an over-dense region of the WHIM in the galaxy filament, or the ionized gaseous halo of a galaxy in the filament (e.g. NGC~5987). In the latter scenario, the absorber could be a fragment accreted by the galaxy from the WHIM. 

(8) Identifying metals associated with BLAs is an efficient way to detect plasma of $10^{4.5} \lesssim T \lesssim 10^6$~K associated with structures that could account for the remaining $\sim 50$\% fraction of the missing baryons in the low-$z$ universe. More such detections of thermally broadened {\Lya} and {\OVI} absorbers are going to be facilitated by the high sensitivity spectroscopy capabilities of the {\it Cosmic Origins Spectrographs} in the far-UV.

\noindent {\bf Acknowledgments :}~The authors thank the STS-125 team for completing a highly successful {\it Hubble Space Telescope} servicing mission in 2009. We are grateful to Gary Ferland and collaborators for developing the Cloudy photoionization code. We thank Orly Gnat for making the computational data on radiatively cooling models public. We also thank an anonymous referee for valuable comments regarding the limitations and uncertainties of the ionization modeling. This research is supported by the NASA {\it Cosmic Origins Spectrograph} program through a sub-contract to the University of Wisconsin-Madison  from the University of Colorado, Boulder. B.P.W acknowledges support from NASA grant NNX-07AH426. This research has made use of the NASA/IPAC Extragalactic Database (NED) which is operated by the Jet Propulsion Laboratory, California Institute of Technology, under contract with the National Aeronautics and Space Administration.

\pagebreak

%%%%%%%%%%%%%%%%%%%%%%%%%%%%%%%%%%%%%%%%%%%%%%%%%%%%%
%%%%%%%%%%%%%%    TABLES    %%%%%%%%%%%%%%%%%%%%%%%%%%%%%%

\clearpage
\begin{deluxetable}{lcccr}
\tabletypesize{\scriptsize} 
\tablewidth{0pt}
\tablecaption{\textsc{COS Observations of MRK~$290$}}
\tablehead{
\colhead{HST ID} &
\colhead{Grating} &
\colhead{Central Wavelength} &
\colhead{Wavelength Range} &
\colhead{Exposure Duration} \\
\colhead{ } &
\colhead{ } &
\colhead{(\AA)} &
\colhead{(\AA)} &
\colhead{(sec)} 
}
\startdata
LB4Q02010 & G130M & 1291 & 1136 -- 1430 & 964 \\
LB4Q02020 & G130M & 1300 & 1146 -- 1440 & 964 \\
LB4Q02030 & G130M & 1309 & 1156 -- 1449 & 964 \\
LB4Q02040 & G130M & 1318 & 1165 -- 1459 & 964 \\
LB4Q02050 & G160M & 1589 & 1400 -- 1761 &1200 \\
LB4Q02060 & G160M & 1600 & 1412 -- 1773 &1200 \\
LB4Q02070 & G160M & 1611 & 1424 -- 1784 &1200 \\
LB4Q02080 & G160M & 1623 & 1436 -- 1796 &1200\\
\enddata
\tablecomments{All observations were carried out on 28 October 2009 as part of cycle 17 COS-GTO proposal ID 1152 (P.I. James Green).} 
\label{tab:tab1}
\end{deluxetable}

%\clearpage
\begin{deluxetable}{lccrrrrr}
\tabletypesize{\scriptsize} 
\tablewidth{0pt}
\tablecaption{\textsc{Line Measurements for the $z(\OVI) = 0.01027$ Absorber}}
\tablehead{
\colhead{Line} &
\colhead{Instrument} &
\colhead{$W_r$} &
\colhead{$v$} &
\colhead{log~$[N~(\cmsq)]$} &
\colhead{$b$} &
\colhead{$[-v, +v]$}  & 
\colhead{Method} \\ 
\colhead{(\AA)} &
\colhead{ } &
\colhead{(m\AA)} &
\colhead{(\kms)} &
\colhead{dex} &
\colhead{(\kms)} &
\colhead{(\kms)} &
\colhead{ }
}
\startdata
{\Lya} & COS/G130M & $...$ &  $2~{\pm}~5$ & $14.35~{\pm}~0.01$ & $55~{\pm}~1$ & $...$ &  Fit \\
{ }	& 		   & $...$ &  $116~{\pm}~5$ & $14.07~{\pm}~0.01$ & $33~{\pm}~1$ & $...$ &   \\
{\Lya} & 		   & $522~{\pm}~10$ & $...$ & $14.30~{\pm}~0.01$ & $64~{\pm}~5$ & [-220, 70] & AOD \\
{ }	&		   & $293~{\pm}~10$ & $...$ & $14.01~{\pm}~0.01$ & $41~{\pm}~4$ & [70, 250] &  \\
{\Lya} & 		   & $808~{\pm}~10$ & $...$ & $14.48~{\pm}~0.01$ & $98~{\pm}~4$ & [-220, 250] & AOD \\
{\Lyg} &  FUSE	   & $< 190$		  & $...$ & $< 15.2$		& $...$	      & [-220, 250] & 3~$\sigma$ \\
\\
\\
\multicolumn{8}{c}{\sc $v(\Lya) \sim 2$ component}
\\
\\
{\OVI}~1032 & FUSE & $...$ & $1~{\pm}~5$ & $13.80~{\pm}~0.05$ & $29~{\pm}~3$ & $...$ & Fit \\
{\OVI}~1032 & FUSE & $74~{\pm}~13$ & $-4~{\pm}~4$ & $13.82~{\pm}~0.08$ & $44~{\pm}~6$ & [-75, 75] & AOD \\
{\OVI}~1032 & FUSE & $49~{\pm}~11$ & $...$ & $13.63~{\pm}~0.08$ & $34~{\pm}~4$ & ... & AOD$^{\dagger}$\\
{\NV}~1239 & COS/G130M & $< 35$ & $...$ & $< 12.9$ & $...$ & [-75, 75] & 3~$\sigma$ \\
{\NV}~1243 & COS/G130M & $< 33$ & $...$ & $< 13.0$ & $...$ & [-75, 75] & 3~$\sigma$ \\
{\CIV}~1548 & COS/G130M & $< 48$ & $...$ & $< 12.9$ & $...$ & [-75, 75] & 3~$\sigma$ \\
{\CIV}~1551 & COS/G130M & $< 43$ & $...$ & $< 13.1$ & $...$ & [-75, 75] & 3~$\sigma$ \\
{\SiIV}~1394 & COS/G130M & $< 42$ & $...$ & $< 12.4$ & $...$ & [-75, 75] & 3~$\sigma$ \\
{\CIII}~977 & COS/G130M & $< 114$ & $...$ & $< 13.3$ & $...$ & [-75, 75] & 3~$\sigma$ \\
{\SiIII}~1207 & COS/G130M & $< 42$ & $...$ & $< 12.1$ & $...$ & [-20, 75] & 3~$\sigma$ \\
{\CII}~1036 & COS/G130M & $< 36$ & $...$ & $< 13.4$ & $...$ & [-75, 75] & 3~$\sigma$ \\
{\CII}~1335 & COS/G130M & $< 43$ & $...$ & $< 13.1$ & $...$ & [-75, 75] & 3~$\sigma$ \\
{\SiII}~1190 & COS/G130M & $< 36$ & $...$ & $< 12.5$ & $...$ & [-75, 75] & 3~$\sigma$ \\
{\FeII}~1145 & COS/G130M & $< 33$ & $...$ & $< 13.5$ & $...$ & [-75, 75] & 3~$\sigma$ \\
\\
\\
\multicolumn{8}{c}{\sc $v(\Lya) \sim 116$ component}
\\
\\
{\OVI}~1038 & FUSE & $< 31$ & $...$ & $< 13.5$ & $...$ & [75, 150] & AOD \\
{\NV}~1239 & COS/G130M & $< 33$ & $...$ & $< 12.4$ & $...$ & [75, 150] & 3~$\sigma$ \\
{\CIV}~1548 & COS/G130M & $< 39$ & $...$ & $< 12.7$ & $...$ & [75, 150] & 3~$\sigma$ \\
{\CIII}~977 & COS/G130M & $26~{\pm}~27$ & $103~{\pm}~7$ & $< 13.2$ & $...$ & [75, 150] & 3~$\sigma$ \\
{\CIII}~977 & COS/G130M & $57~{\pm}~22$ & $103~{\pm}~7$ & $13.11~{\pm}~0.22$ & $...$ & [80, 130] & 3~$\sigma$ \\
{\SiIII}~1207 & COS/G130M & $< 37$ & $...$ & $< 12.0$ & $...$ & [75, 150] & 3~$\sigma$ \\
{\CII}~1335 & COS/G130M & $< 40$ & $...$ & $< 13.0$ & $...$ & [75, 150] & 3~$\sigma$ \\
\enddata
\tablecomments{Column 3 lists the rest-frame equivalent width,  column 4 gives the velocity centroid of line components derived from profile fitting, column 5 lists the column density, column 6 is the Doppler parameter and column 7 the integration velocity range to derive the respective $W_r$, $N$ and $b$. 
\\
\\
All lines other than {\Lya} and {\OVI} are non-detections at the 3~$\sigma$ significance level. For the two components in {\Lya}, different velocity ranges were chosen for measurements of the metal ions. The {\OVIdblt}~{\AA}, {\Lyg} and {\CIII}~$\lambda 977$~{\AA}  are from {\it FUSE}. 
\\
\\
For {\Lya} and {\OVI}~$\lambda 1032$~{\AA}, we list the line parameters derived from fitting a Voigt function to the profile and also using the apparent optical depth (AOD) method of Sembach \& Savage (1992). The profile fit was performed using the Fitzpatrick \& Spitzer (1997) routine with the fit model for {\Lya} convolved with the COS/G130M line spread function described in Ghavamian {\etal}(2009).  The velocity errors listed are dominated by the velocity calibration errors of $\sim~{\pm}~5$~{\kms} for both COS and $FUSE$. For the $FUSE$ {\OVI}~$\lambda 1032$~{\AA} line a Gaussian kernel of FWHM = 25~{\kms} was convolved with the Voigt model to fit the observed profile.
\\
\\
$^{\dagger}$ The measurement from \citet{wakker09}.} 
\label{tab:tab2}
\end{deluxetable}

\clearpage
%\begin{landscape}
%\begin{sidewaystable}
\begin{deluxetable}{lcccccccc}
\tabletypesize{\scriptsize} 
\tablewidth{0pt}
\tablecaption{\textsc{Broad {\Lya} Absorbers with Metals}}
\tablehead{
\colhead{Target} &
\colhead{$z$} &
\colhead{log~$[N(\HI)~{\cmsq}]$} &
\colhead{$b(\HI)$} &
\colhead{log~$[N(\OVI)~{\cmsq}]$} &
\colhead{$b(\OVI)$} &
\colhead{$|\Delta v|$} &
\colhead{$T$} &
\colhead{Reference} \\
\colhead{} &
\colhead{} &
\colhead{(dex)} &
\colhead{(\kms)} &
\colhead{(dex)} &
\colhead{(\kms)} &
\colhead{(\kms)} &
\colhead{(K)} &
\colhead{}
}
\startdata
3C~273 & 0.00337 & $14.27~{\pm}~0.08$ & $53~{\pm}~3$ & $13.24~{\pm}~0.21$ & $34~{\pm}~6$ & $2$ & $1.1 \times 10^5$ & $1$ \\
MRC~2251-178 & $0.00755$ & $13.43~{\pm}~0.05$ & $64~{\pm}~4$ & $13.87~{\pm}~0.26$ & $35~{\pm}~5$ & $8$ & $1.9 \times 10^5$ & $1$ \\
MRK~876 & $0.00312$ & $14.27~{\pm}~0.08$ & $77~{\pm}~2$ & $13.18~{\pm}~0.21$ & $29~{\pm}~3$ & $9$ & $3.3 \times 10^5$ & $1$ \\
PG~1444+407 & $0.22032$ & $13.65~{\pm}~0.05$ & $86~{\pm}~15$ & $13.94~{\pm}~0.07$ & $36~{\pm}~8$ & $3$ & $3.9 \times 10^5$ & $2$ \\
H~1821+643 & 0.12147 & $13.78~{\pm}~0.17$ & $85^{+37}_{-26}$ & $14.02~{\pm}~0.07$ & $59^{+30}_{-20}$ & $10$ & $2.4 \times 10^5$ & $2$ \\
H~$1821+643$ & $0.26656$ & $13.63~{\pm}~0.02$ & $44~{\pm}~2$ & $13.63~{\pm}~0.04$ & $25~{\pm}~3$ & $6$ & $0.8 \times 10^5$ & $2$ \\
PG~1259+593 & $0.31978$ & $13.98~{\pm}~0.06$ & $74~{\pm}~9$ & $13.49~{\pm}~0.07$ & $19~{\pm}~4$ & $10$ & $3.3 \times 10^5$ & $3$ \\
\\
\\
& & & & log~$[N(\CIII)~{\cmsq}]$ & $b(\CIII)$ & & \\
\\
\\
H~$1821+643$ & $0.22638$ & $13.50~{\pm}~0.01$ & $51~{\pm}~2$ & $12.56~{\pm}~0.05$ & $28~{\pm}~4$ & $1$ & $1.2 \times 10^5$ & $4$
\\
\enddata
\tablecomments{A listing of known BLAs with metals with absorption aligned within $|\Delta v| = 15$~{\kms} in velocity. The references are (1) \citet{wakker09} (2) \citet{tripp08} (3) Richter {\etal}(2004) (4) Narayanan, Savage \& Wakker (2010). Column 1 is the sight line, column 2 the redshift of the absorber, column 3 and 5 are the column densities of {\HI} and metal ion, column 4 and 6 are their respective $b$-values, column 7 is the difference in velocity between the {\Lya} and the metal-line, column 8 is the temperature derived using the $b$-values for {\HI} and the metal ion assuming that the two species live in the same gas phase and the listed $b$ is a quadrature sum of thermal and non-thermal $b$ components.}
\label{tab:tab3}
\end{deluxetable}
%\end{landscape}

%%%%%%%%%%%%%%%%%%%%%%%%%%%%%%%%%%%%%%%%%%%%%%%%%%%%%
%%%%%%%%%%%%%%    FIGURES    %%%%%%%%%%%%%%%%%%%%%%%%%%%%%%

\clearpage
\begin{figure*}
\begin{center}
\includegraphics[scale=0.8]{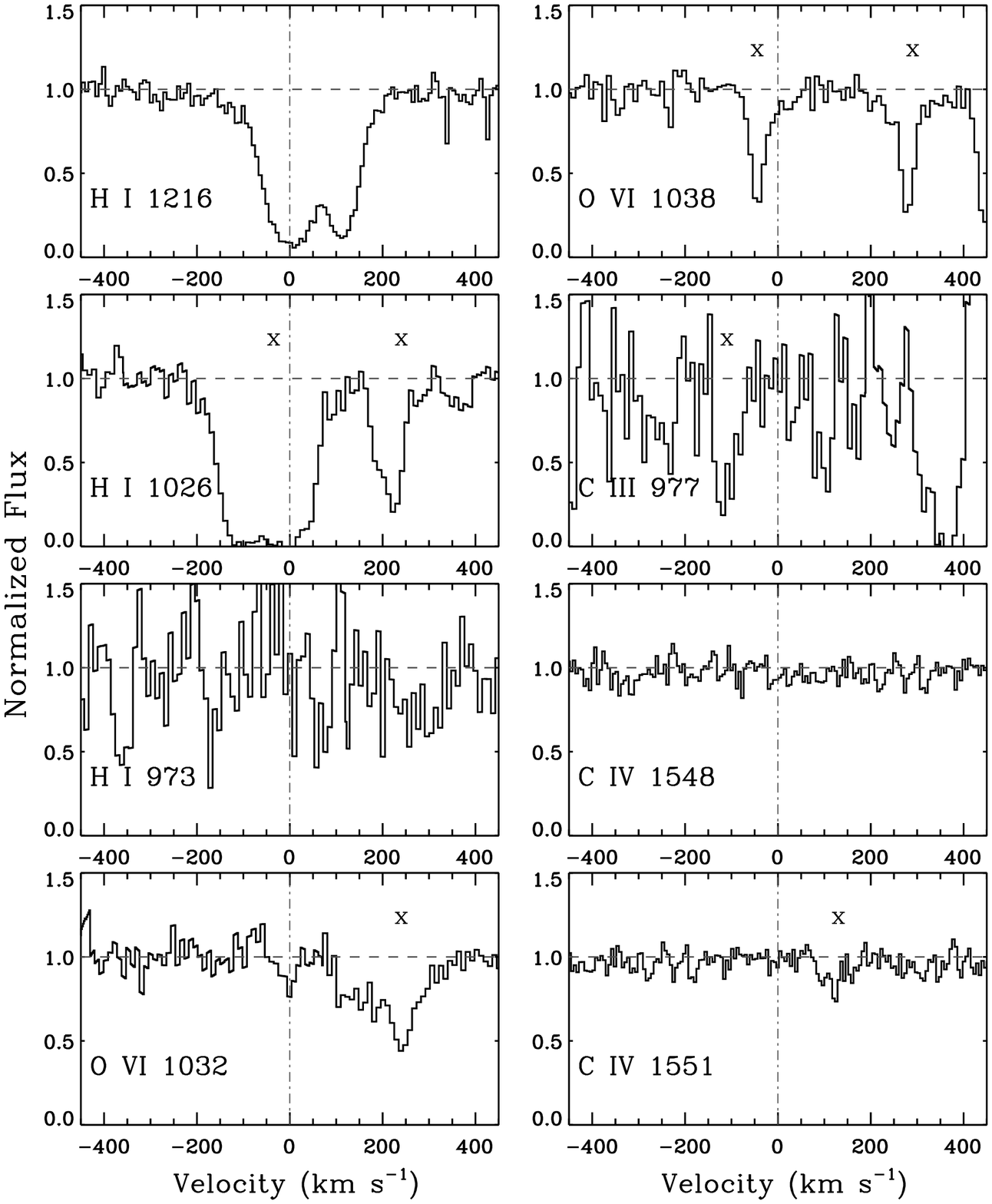}
\end{center}
\protect
{\large{Fig.~1a.~---~Continuum normalized spectrum of Mrk~$290$ showing absorption profiles of {\Lya}, and {\OVI} and the wavelength regions of other prominent lines in the $z(\OVI) = 0.01027$ rest-frame of the absorption system. The {\OVIdblt}~{\AA}, {\Lyg} and {\CII}~$\lambda 1036$~{\AA}, {\Lyb} and {\Lyg} are lower $S/N$ $FUSE$ spectra \citep{wakker09} and the rest are $HST$/COS G130M and G160M observations. Every other metal line, except {\OVI}, is a non-detection at the $3~\sigma$ significance level. The line measurements are listed in Table 2. {\Lyb} and {\OVI}~$\lambda 1038$~{\AA} are strongly contaminated by Galactic {\CII}~$\lambda 1036$~{\AA} and {\ArI}~$\lambda 1048$~{\AA} respectively. The wavelength region corresponding to the {\Lyg} does not suggest any significant unresolved saturation in {\HI}. Features which are not part of the absorption system are labeled with "x" in the respective panels.}}
\label{fig:1a}
\end{figure*}

\clearpage
\begin{figure*}
\begin{center}
\includegraphics[scale=0.9]{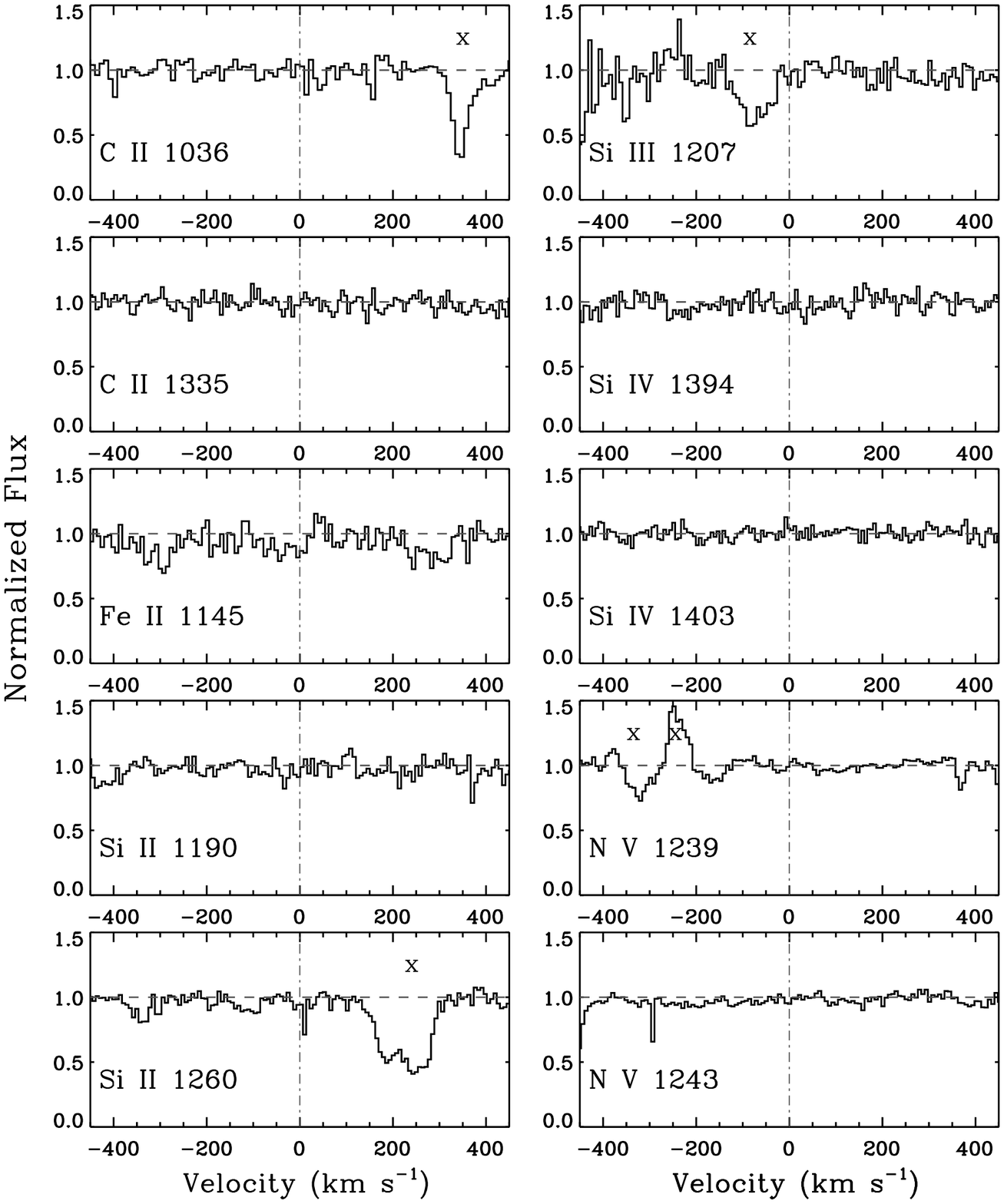}
\end{center}
\protect
{Fig.~1b.~---~Continuation of Figure 1a.}
\label{fig:1b}
\end{figure*}

\setcounter{figure}{1}
\clearpage
%\begin{landscape}
\begin{figure*}
\begin{center}
\includegraphics[scale=0.7]{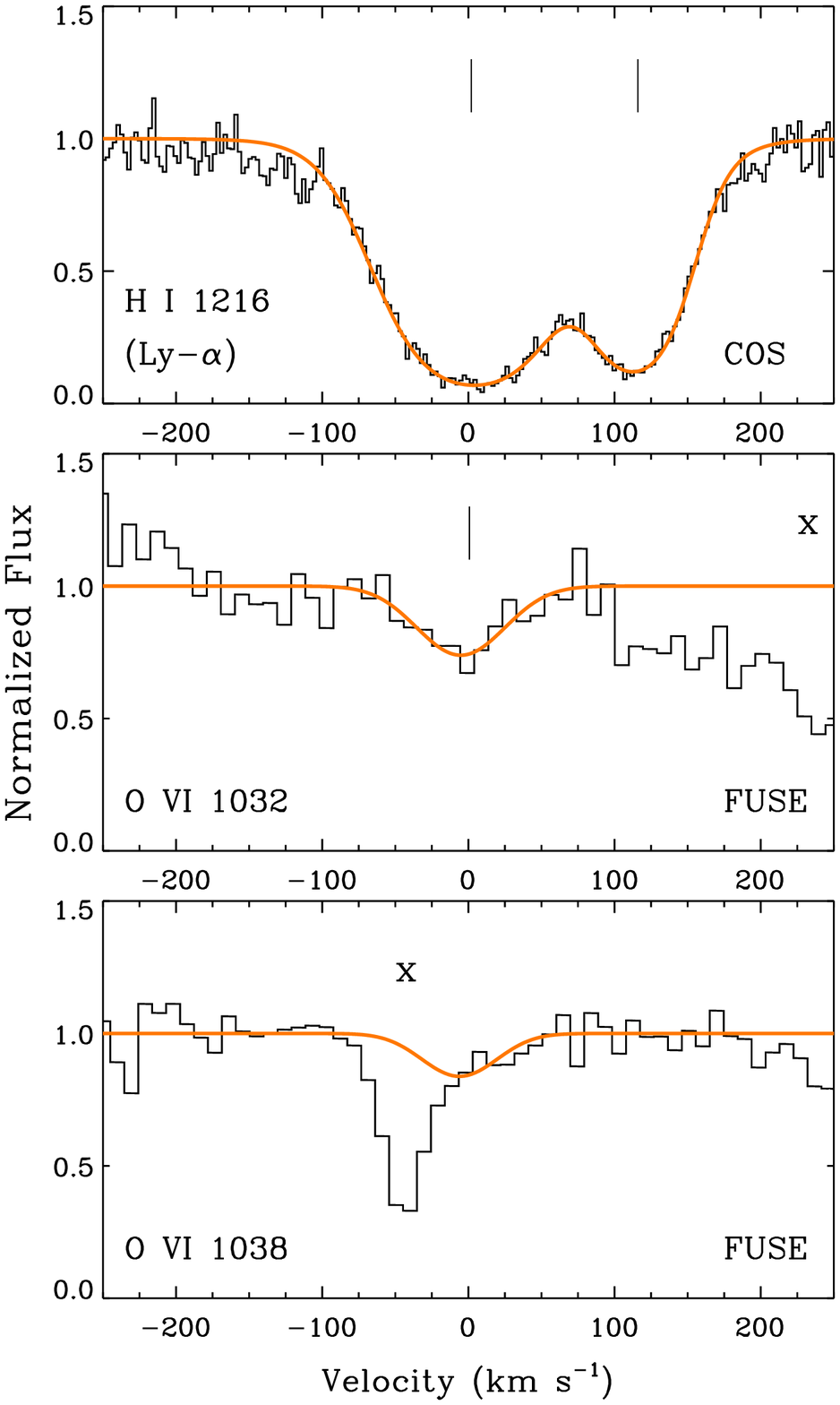}
\end{center}
\protect
\caption{\large{{\it Top Panel :}~Shows the two component Voigt profile model superimposed on the {\Lya} absorption. The vertical axis is continuum normalized flux and the horizontal axis is velocity in the rest-frame of the absorber, where $v = 0$~{\kms} corresponds to $z(\OVI) = 0.01027$. Features which are not part of the absorption system are marked using "x" in the respective panels. The tick marks above the {\Lya} profile show the centroids of the two absorbing components at $v = 2~{\pm}~5$~{\kms} and $v = 116~{\pm}~5$~{\kms} respectively. The model Voigt profiles were convolved with the COS/G130M line spread function given by Ghavamian {\etal}(2009). {\it Middle \& Bottom Panels :}~Show the fit to the {\OVI}~$\lambda 1032$~{\AA} and the contaminated {\OVI}~$\lambda 1038$~{\AA} lines respectively. A Gaussian of FWHM = 25~{\kms} corresponding to the resolution of $FUSE$ was used to derive the model fit. The fit parameters are given in Table 2.}}
\label{fig:2}
\end{figure*}
%\end{landscape}

\clearpage
\begin{figure*}
\begin{center}
\epsscale{0.9}
\plotone{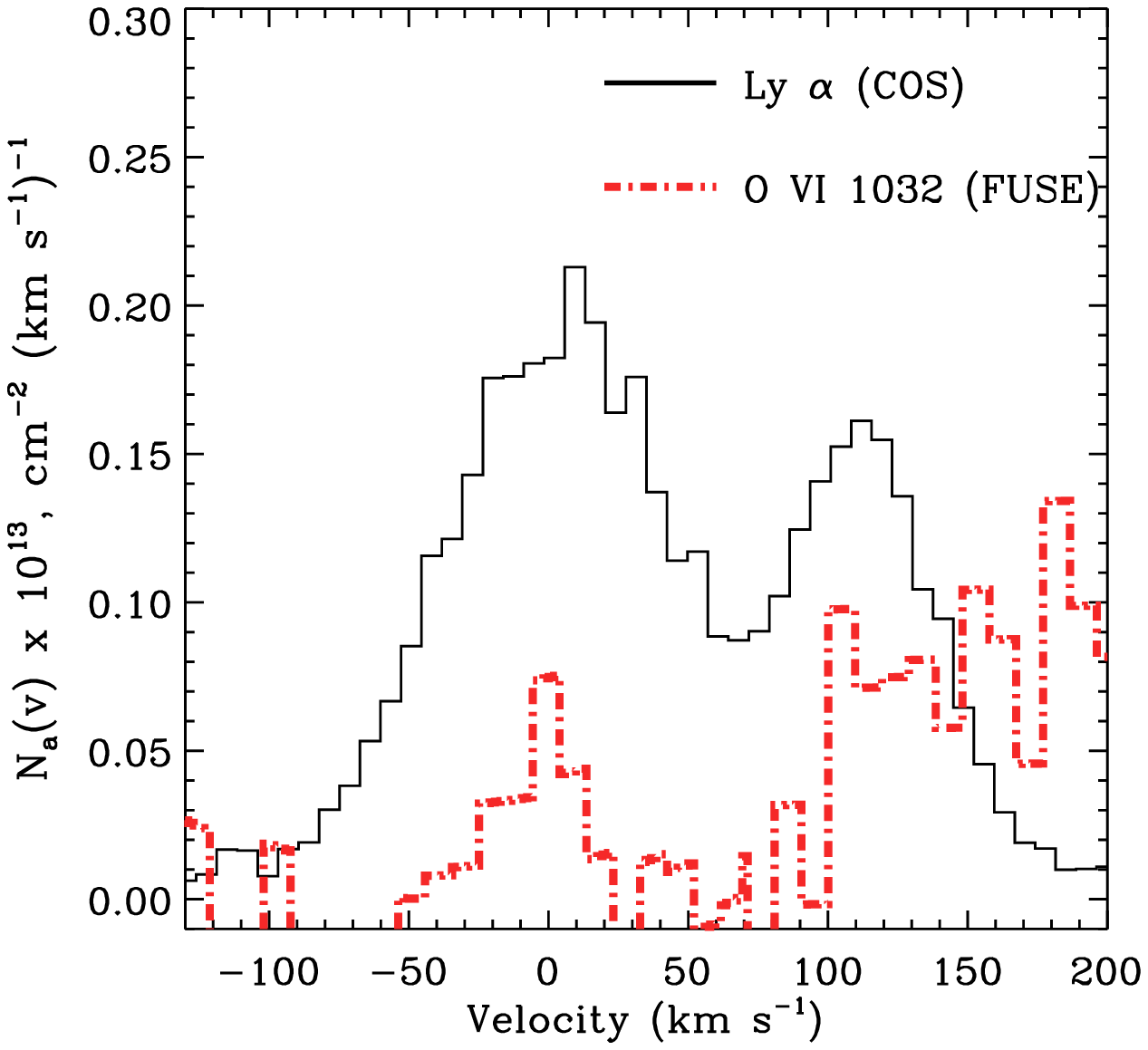}
\end{center}
\protect
\caption{\large{Apparent column density ($N_a(v)$) profile plots for {\Lya} and {\OVI}~$\lambda 1032$~{\AA}.  The {\OVI}~$\lambda 1032$~{\AA} absorption at $v = 1~{\pm}~5$~{\kms} is kinematically well-aligned with the BLA component of the {\HI} absorption at $v = 2~{\pm}~5$~{\kms}, suggesting a single-phase origin for {\HI} and {\OVI}. The {\Lya} and {\OVI} are binned to approximately the same resolution of $\sim 9$~{\kms} velocity bins. The $N_a(v)$ for {\OVI}~$\lambda 1032$ is contaminated at $v \geq 100$~{\kms} by a feature unrelated to this system.}}
\label{fig:3}
\end{figure*}

\clearpage
%\begin{landscape}
\begin{figure*}
%\begin{sidewaysfigure}
\begin{center}
\includegraphics[scale=0.6,angle=270]{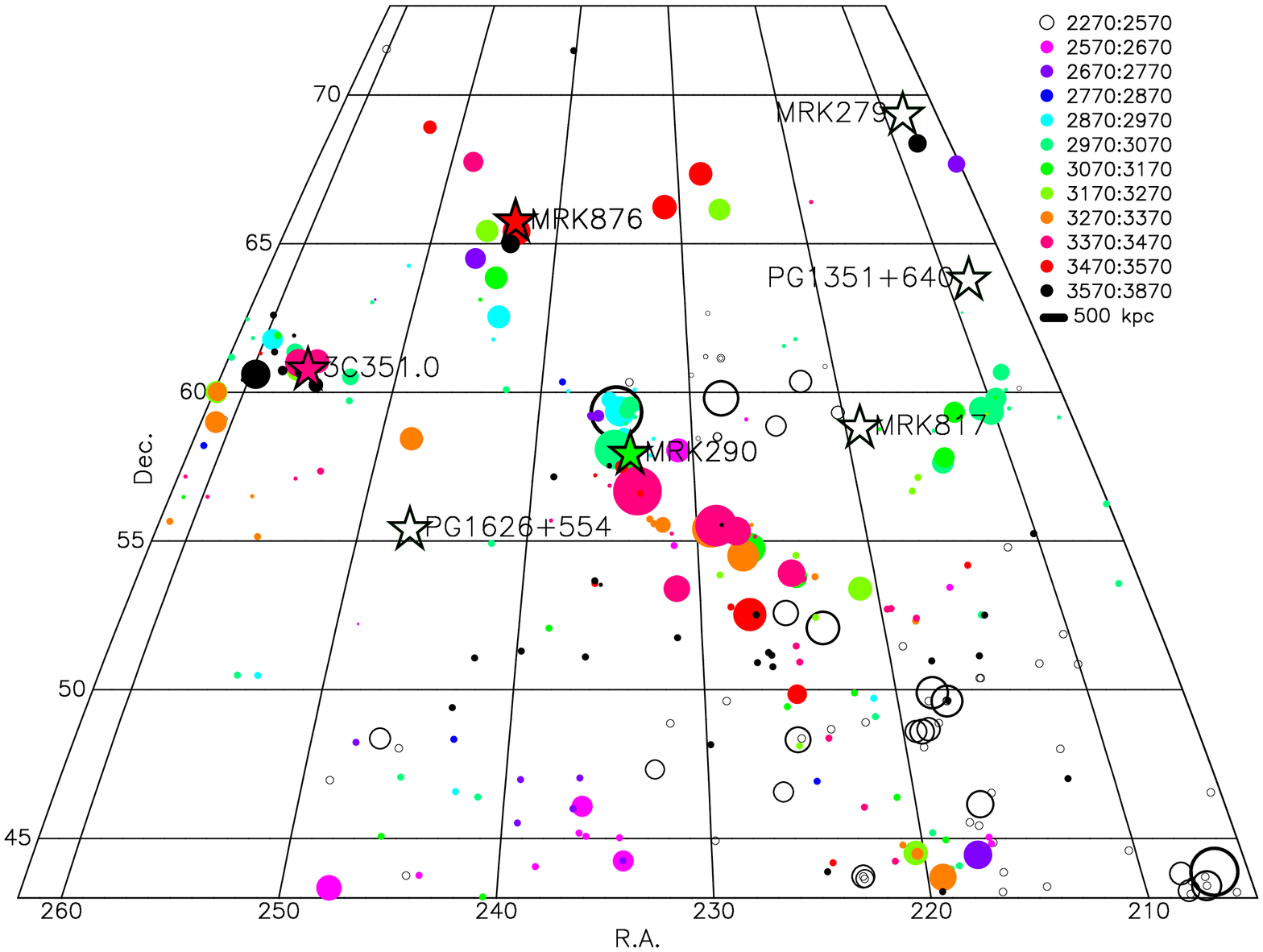}
\includegraphics[scale=0.6,angle=270]{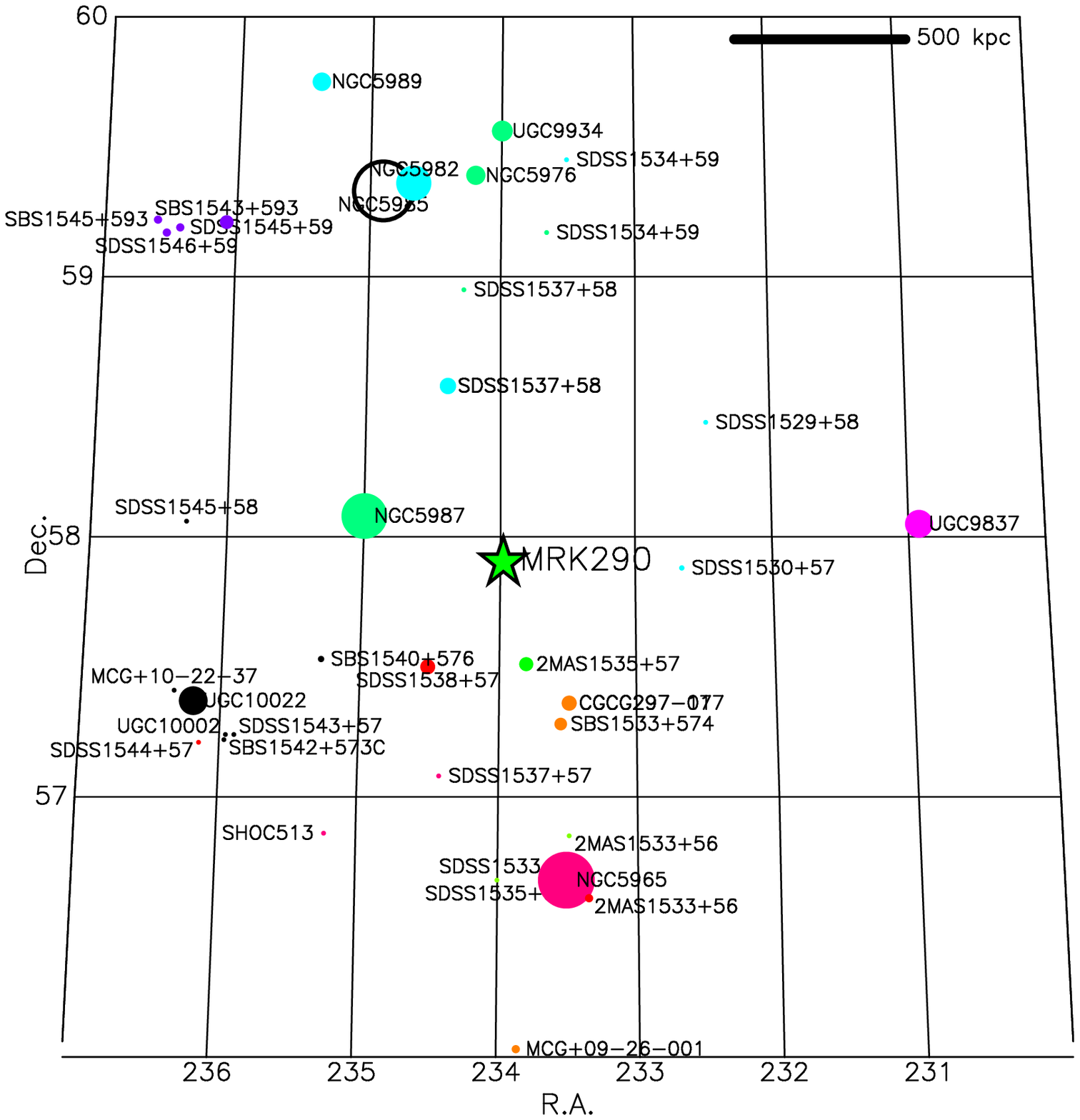}
\end{center}
\protect
\small{Fig.~4a.~---~{\it Top Panel :}~Galaxies in the foreground field of Mrk~$290$ within a systemic velocity of $v \sim 1000$~{\kms} and projected distance of $\sim 4~h^{-1}$~Mpc from the BLA-{\OVI} absorber. The {\it circles} represent galaxies and the {\it stars} represent sight-lines towards AGNs in the field. The distribution pattern of the galaxies shows that the sight line to Mrk 290 is intercepting a megaparsec scale filament of galaxies at the location of the BLA - {\OVI} absorber. Those AGNs with an intervening absorption system identified at the redshift of the galaxy filament are plotted in {\it filled star} symbols. The size of the {\it circles} are proportional to the size of the galaxies given by their $D_{25}$ major diameter. The color coding of the symbols represent various heliocentric velocity ranges. The information on galaxies and AGNs in this region of the sky were retrieved from NED. {\it Bottom Panel :} A zoom-in view of the $\sim 1~h^{-1}$~kpc region centered around Mrk~290. The galaxy IDs are labeled. Among the extended sources, NGC~5987 is closest in impact parameter ($\rho = 424~h^{-1}$~kpc) and velocity ($|\Delta v| = 73$~{\kms}) to the BLA - {\OVI} absorber.} 
\label{fig:4a}
%\end{sidewaysfigure}
\end{figure*}
%\end{landscape}

\clearpage
\begin{figure*}
\begin{center}
\epsscale{0.9}
%\plotone{f4b.eps}
{\Large{Fig.~4b.---~~A Digital Sky Server (DSS) image of NGC~5987 with the line of sight to Mrk 290 marked. Figure not included in the astro-ph version due to large file size.}}
\end{center}
\protect
%%{\large{Fig.~4b.~---~~A Digital Sky Server (DSS) image of NGC~5987 rotated such that its major axis is horizontal. The X and Y axes are projected distances in units of kiloparsec from the galaxy. The sight-line towards Mrk~290 is indicated by the open {\it plus} symbol. Among the several galaxies along the megaparsec scale filament displayed in Figure 4a, NGC~5987 is at the smallest impact parameter ($\rho = 424~h^{-1}$~kpc) and systemic velocity ($|\Delta v| = 73$~{\kms}) from the BLA - {\OVI} absorber at $z = 0.01027$. Its heliocentric velocity, luminosity distance and major diameter at the 25.0 B-mag arcsec$^2$ isophote (RC3 catalogue's standard diameter system) are also labeled. NGC~5987 is a $\sim 2L^*$ galaxy of Sb morphology with an apparent B-band magnitude of $B = 12.72$ \citep{devaucouleurs91}. The galaxy is oriented almost edge-on with respect to the line of sight.}}
\label{fig:4b}
\end{figure*}

\setcounter{figure}{4}
\clearpage
\begin{figure*}
\begin{center}
\includegraphics[scale=0.9,angle=90]{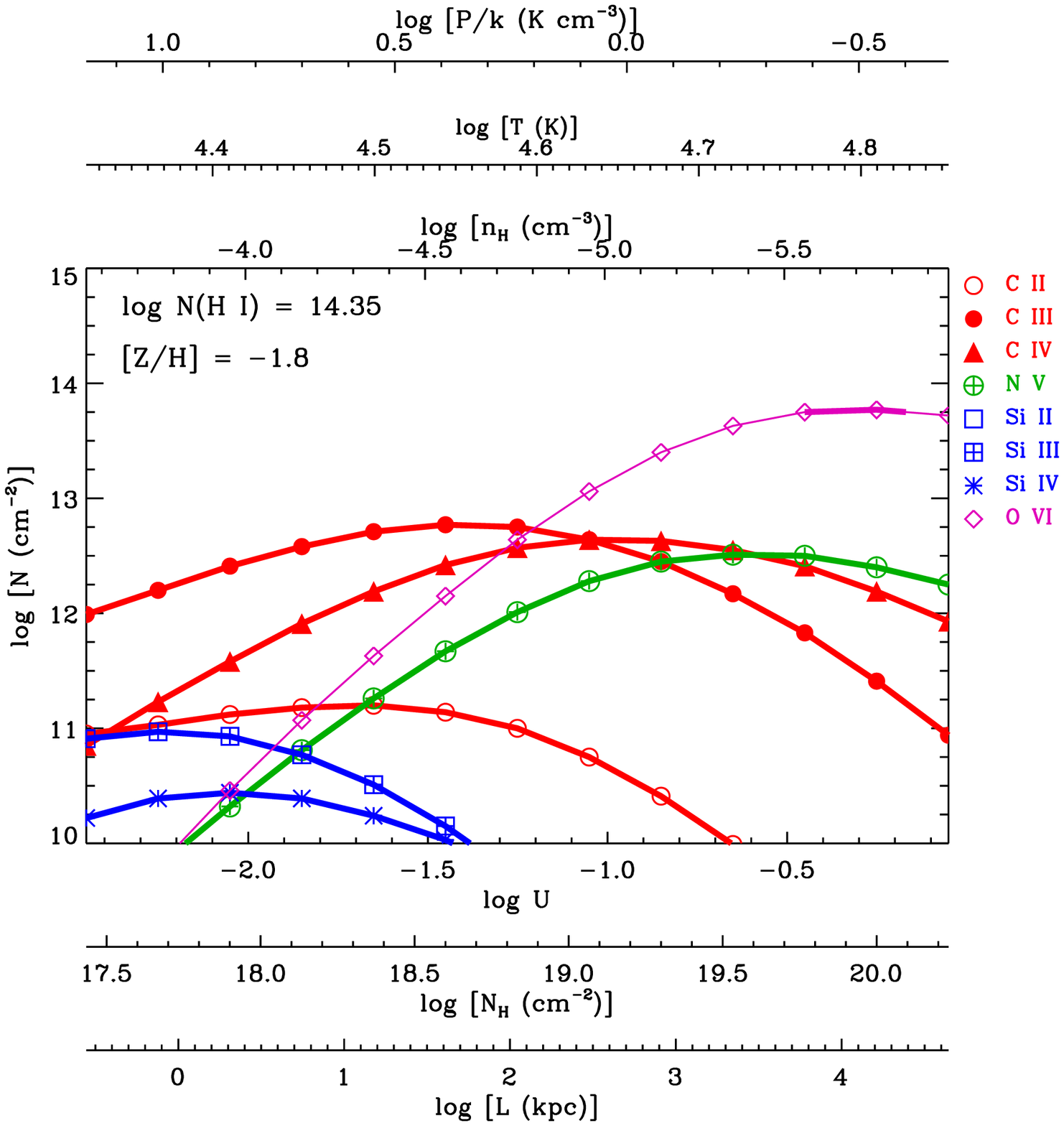}
\end{center}
\protect
\caption{\large{Photoionization model predictions of column densities for the BLA - {\OVI} absorber. The $N(\HI) = 10^{14.35}$~{\cmsq} is the column density for the $v = 2$~{\kms} BLA component of the {\HI} absorption. The ionizing background is from the Haardt \& Madau (2001) model for $z = 0.01027$ and includes UV photons from quasars and young star forming galaxies. The acceptable range of column densities for each ion, based on measurement, are highlighted in the photoionization curves using {\it thick} lines. Except for {\OVI}, all other ions are non-detections at the $3~\sigma$ significance level. The column density measurements are therefore upper limits for those ions. The multiple X-axes are the ionization parameter (log~$U$), total hydrogen column density ($N(\H)$,~{\cmsq}), hydrogen number density ($n_{\H}$,~{\cc}), temperature ($T$, K), gas pressure ($p/K$,~{\cc}~K, and size of the absorbing region ($L = N(\H)/n_{\H}$,~cm). The models assume solar relative elemental abundances of Asplund {\etal}(2009). The metallicity for which the models are able to recover the measured $N(\OVI)$ is -1.8 dex of solar.}}
\label{fig:5}
\end{figure*}

\clearpage
\begin{figure*}
\begin{center}
\includegraphics[scale=0.9,angle=90]{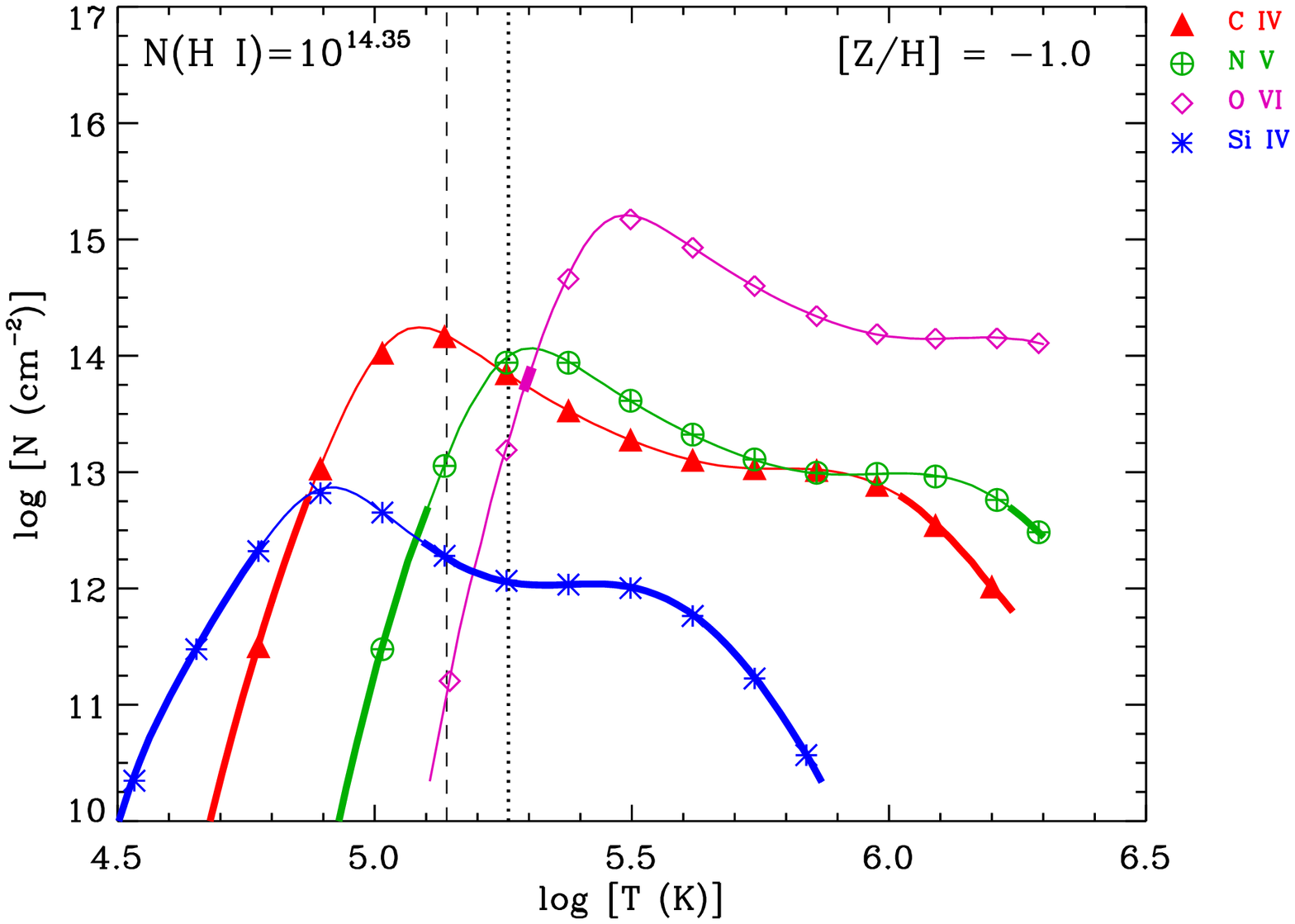}
\end{center}
\protect
\caption{\large{Collisional ionization equilibrium curves for the high ions in the BLA - {\OVI} absorber based on the models of Gnat \& Sternberg (2007). The {\HI} column density in these models are set to the measured value of $10^{14.35}$~{\cmsq} (see table 2). We use the updated solar elemental abundances of \citet{asplund09}. Except {\OVI}, the other high ions are non-detections at $\geq 3~\sigma$ significance. Their column densities are therefore upper limits. The region of the {\OVI} CIE curve where the measured $N(\OVI)~{\pm}~\sigma[N(\OVI)]$ is recovered is drawn in {\it thick}. Similarly, the regions of the {\CIV}, {\NV} and {\SiIV} CIE curves that are consistent with the measured upper limits of their column densities are drawn using {\it thick} lines. The vertical {\it dashed} line marks $T = 1.4 \times 10^5$~K, the temperature given by the combined BLA and {\OVI} $b$-values. The vertical {\it dotted} line corresponds to the temperature upper limit of $T = 1.8 \times 10^5$~K if the entire $b(\HI) = 55$~{\kms} line width for the BLA is thermal.}}
\label{fig:6}
\end{figure*}

\clearpage
\begin{figure*}
\begin{center}
\includegraphics[scale=0.9,angle=90]{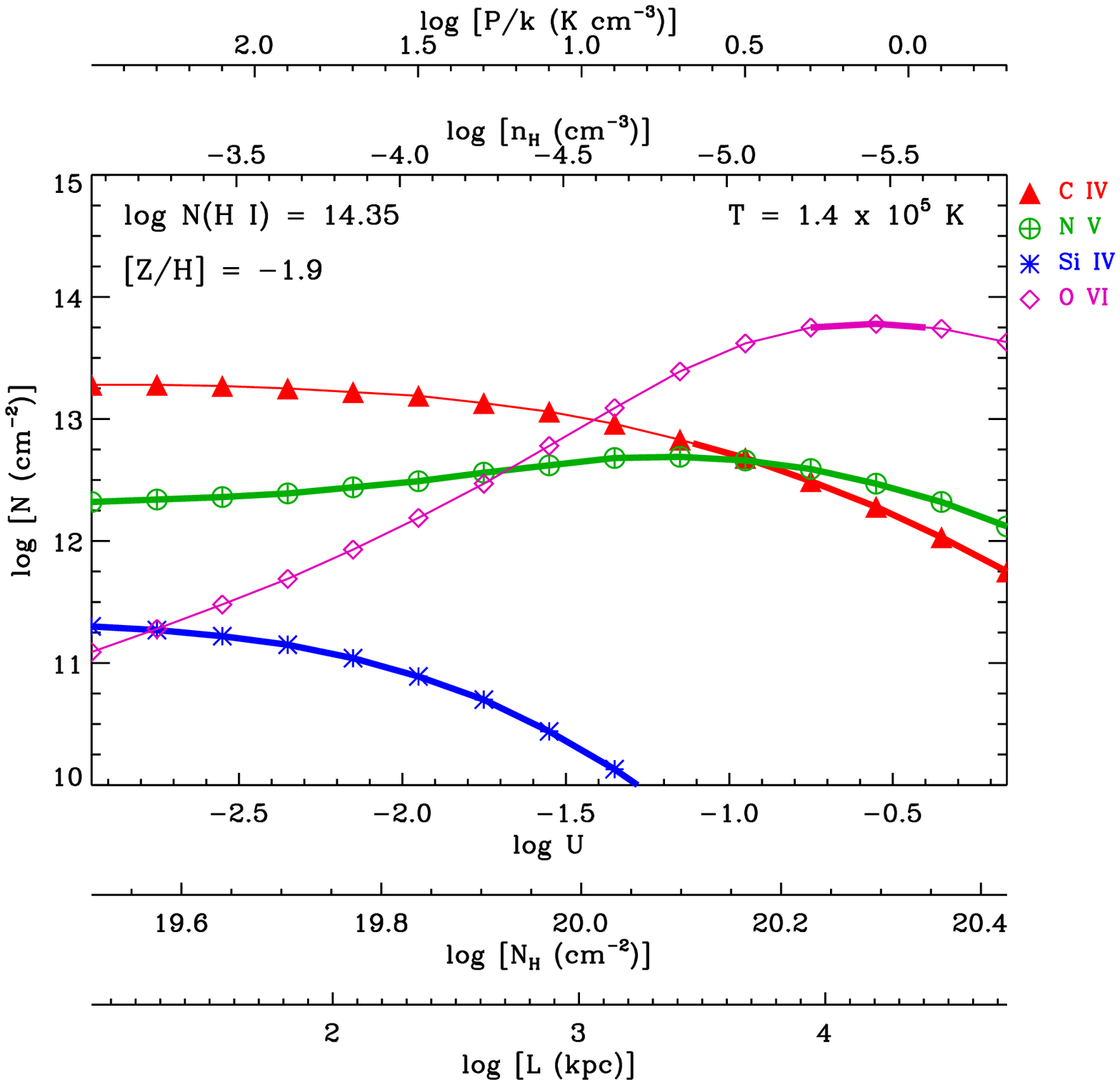}
\end{center}
\protect
\caption{\large{Column density predictions from {\it hybrid} models where both collisional and photoionization reactions are allowed to contribute to ion fractions. The models were computed using Cloudy [ver.C08.00, \citet{ferland98}]. The temperature in the models is fixed to $T = 1.4 \times 10^5$~K derived for the BLA - {\OVI} phase, and the $N(\HI) = 10^{14.35}$~{\cmsq} measured from profile fit to {\Lya} (see table 2). The acceptable column density ranges based on measurements for the various ions are draw in {\it thick}. The observed $N(\OVI)$ is recovered at log~$U \sim -0.6$ for a metallicity of -1.9~dex, assuming solar elemental abundances of \citet{asplund09}.}}
\label{fig:7}
\end{figure*}

\clearpage
\begin{figure*}
\begin{center}
\includegraphics[scale=0.9,angle=90]{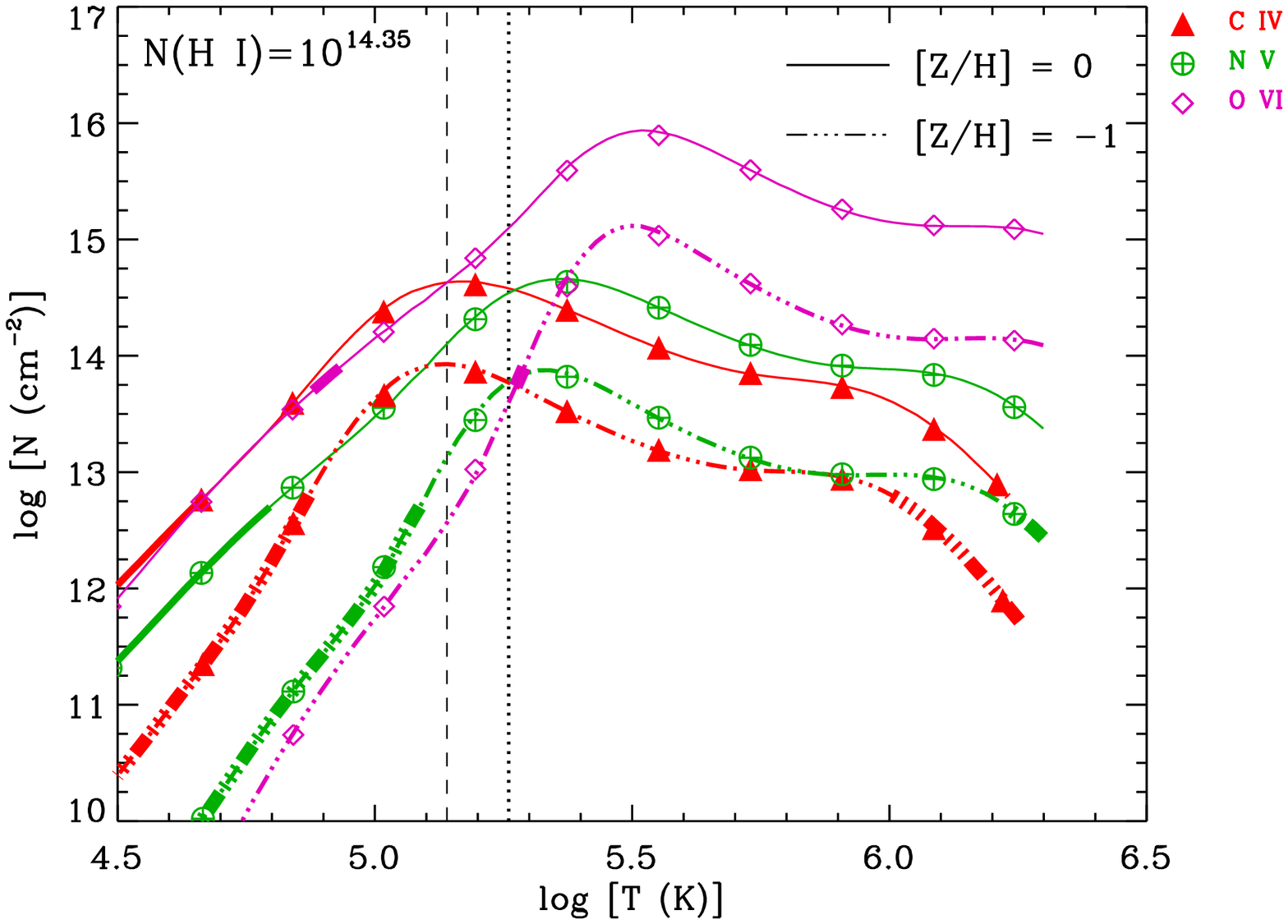} 
\end{center}
\protect
\caption{\large{Nonequilibrium collisional ionization models of Gnat \& Sternberg (2007) for 1/10 solar ({\it dash - dot} line) and solar ({\it solid} line) metallicities and updated solar relative elemental abundances of \citet{asplund09}. The acceptable column density ranges for the various ions are drawn in {\it thick} on each ion curve. Due to rapid rate of cooling, which is higher for the higher gas metallicities, the ion fractions of {\OVI} (and other high ions) are greater than the predictions of CIE in the temperature range $10^{4.5} \leq T \leq 10^{5.5}$~K. The vertical {\it dashed} line marks $T = 1.4 \times 10^5$~K, the temperature given by the combined BLA and {\OVI} $b$-values. The vertical {\it dotted} line corresponds to the temperature upper limit of $T = 1.8 \times 10^5$~K if the entire $b(\HI) = 55$~{\kms} line width for the BLA is thermal.}}
\label{fig:8}
\end{figure*}

\clearpage
%\begin{sidewaysfigure}
%\begin{landscape}
\begin{figure*}
\begin{center}
\includegraphics[scale=0.7,angle=90]{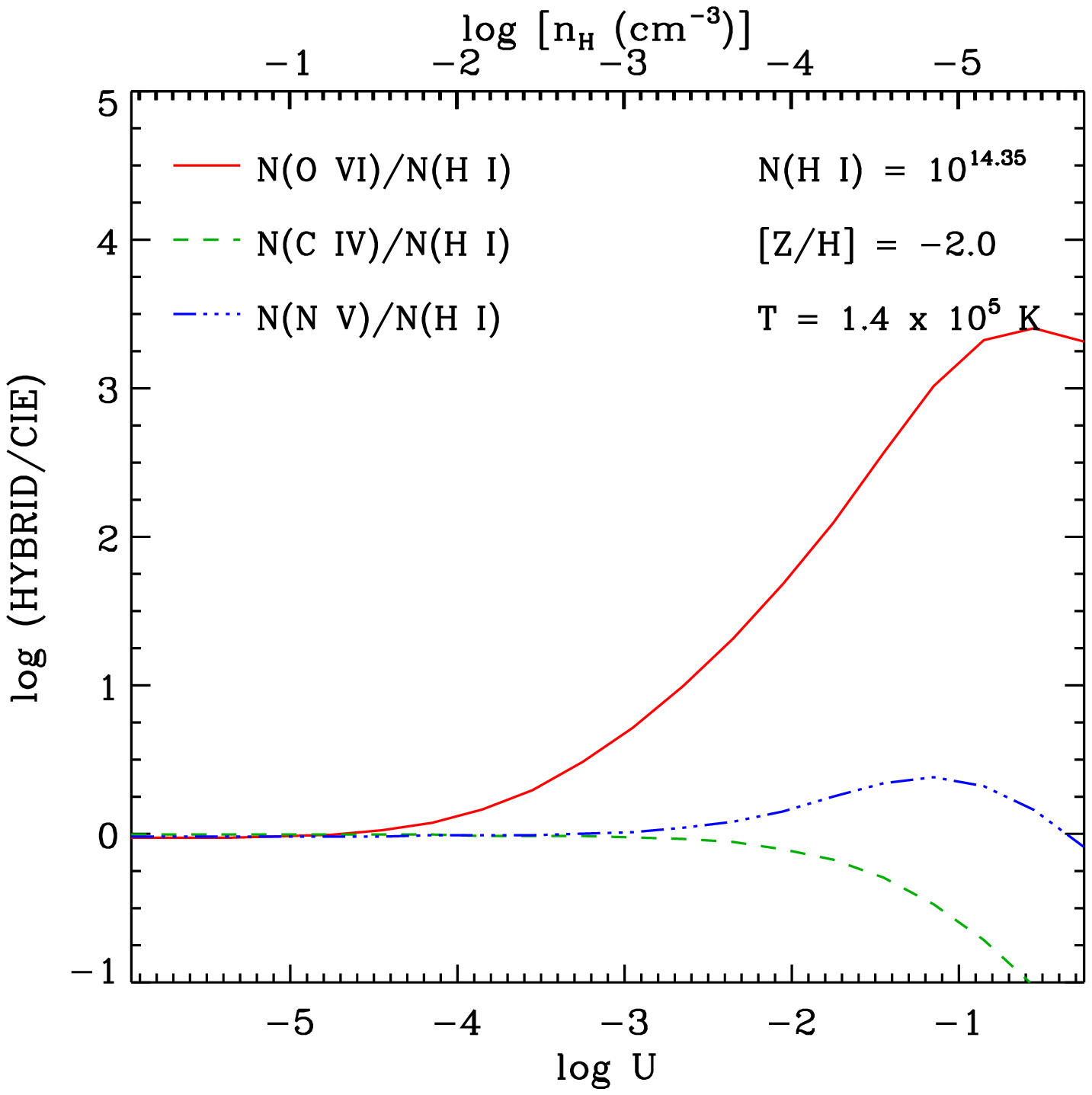}
\includegraphics[scale=0.7,angle=90]{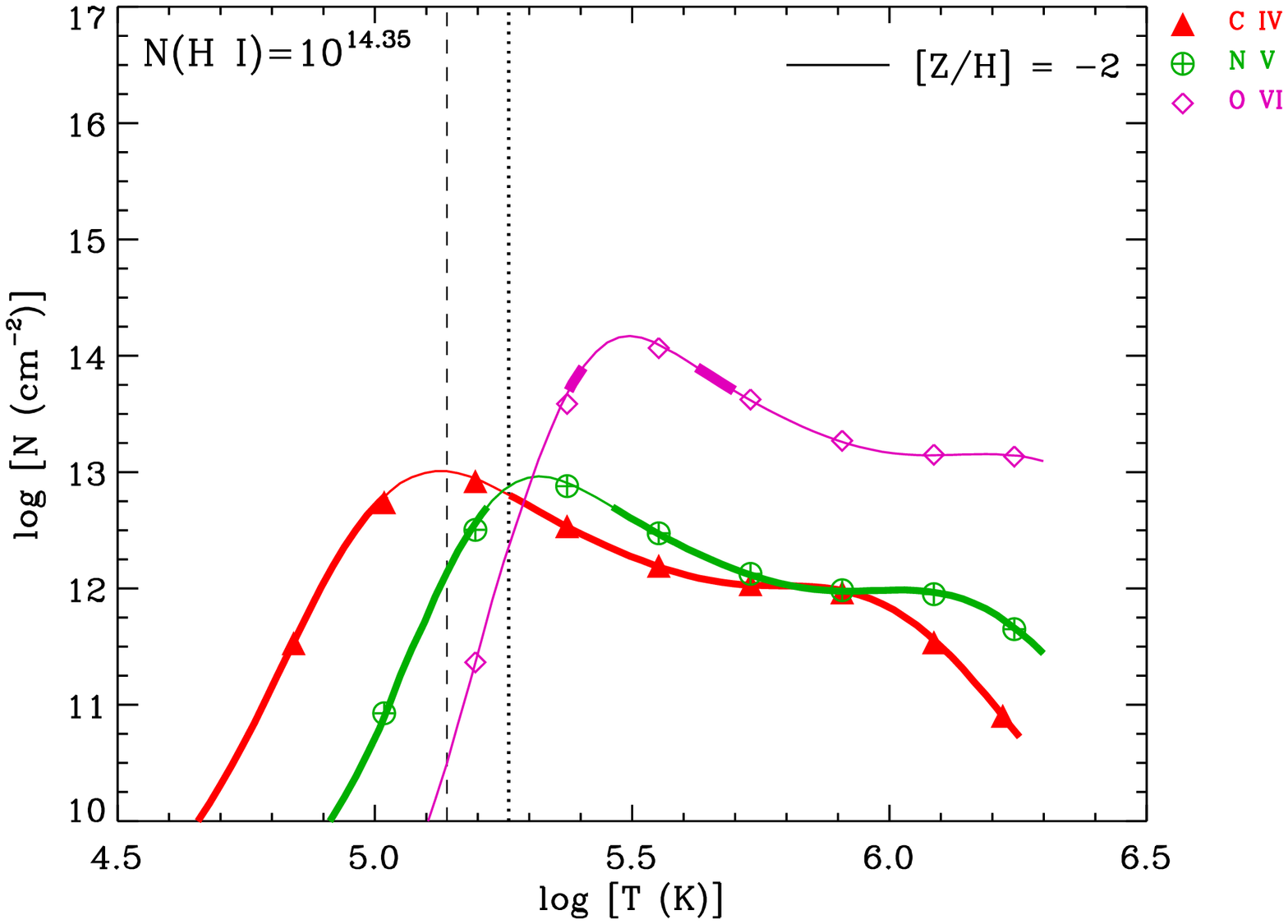}
\end{center}
\protect
\caption{\large{{\it Top Panel :}~CIE - photoionization {\it hybrid} models computed using Cloudy [ver.C08.00, \citet{ferland98}] for solar metallicity. The Y-axis shows the ion ratio predictions in the {\it hybrid} models normalized to their corresponding value for CIE. {\it Bottom Panel :}~Column density predictions made by noneq-CI models of Gnat \& Sternberg(2007) for a radiatively cooling gas at different temperatures and a metallicity of 1/100 solar. The vertical {\it dashed} line marks $T = 1.4 \times 10^5$~K, the temperature given by the combined BLA and {\OVI} $b$-values. The vertical {\it dotted} line corresponds to the temperature upper limit of $T = 1.8 \times 10^5$~K if the entire $b(\HI) = 55$~{\kms} line width for the BLA is thermal. To approximately allow for photoionization, the model predictions in the {\it bottom panel} at $T = 1.4 \times 10^5$~K need to be corrected by the values for each ion given in the {\it left panel} for the appropriate value of log~$U$.}}
\label{fig:9}
%\end{sidewaysfigure}
\end{figure*}
%\end{landscape}

\end{document}